*CORRESPONDENCE
Masahiro Fujita
m.fujita@kansai-u.ac.jp

# Psychological features of dispute content and public acceptance of AI in legal adjudication: evidence for systematic variation beyond individual differences

Masahiro Fujita[1]* and Eiichiro Watamura[2]

[1]Faculty of Sociology, Kansai University, Suita, Japan, [2]Graduate School of Human Sciences, Osaka University, Suita, Japan

Public acceptance of artificial intelligence (AI) in legal decision-making has been primarily explained through individual differences in personality traits and general attitudes toward technology. However, emerging evidence suggests that contextual features of legal disputes themselves may systematically influence preferences for AI versus human adjudicators. Across two studies with Japanese participants ($N$ = 1,384 and $N$ = 596), we examined whether psychological characteristics of dispute content—beyond demographics and individual traits—shape acceptability judgments for algorithmic adjudication. Study 1 employed exploratory factor analysis on acceptability ratings across 46 legal dispute vignettes, revealing a robust two-dimensional structure distinguishing interpersonal-relational disputes (where human adjudicators were strongly preferred) from institutional-procedural disputes (where AI acceptance was comparatively higher, though not surpassing human preference in most cases). Study 2 replicated this dimensional structure in an independent sample and demonstrated that experimentally manipulated contextual features—emotional involvement and prototypicality—systematically modulated acceptability judgments, with effects varying by dispositional trust, AI-specific attitudes, and gender. AI-specific expectations emerged as the strongest predictor of acceptance ($\eta^2$ = 0.252), and a three-way interaction among emotional involvement, gender, and prototypicality indicated that contextual effects are moderated by individual characteristics. These findings suggest that the psychological features of dispute content constitute an overlooked dimension in AI acceptance research, extending beyond technology acceptance models to fundamental questions about how individuals construe social problems and allocate adjudicative authority. We discuss limitations related to measurement approaches, alternative psychological mechanisms, and directions for future research employing real-world case materials and direct assessment of cognitive processes.

# 1 Introduction

## 1.1 AI integration in judicial decision-making: current state and challenges

AI technologies are increasingly deployed in judicial contexts for case management, outcome prediction, and dispute resolution (Aletras et al., 2016; Ahmad, 2025). While technically feasible, successful implementation depends critically on social acceptance. Democratic legitimacy requires technological sophistication to align with citizens' normative expectations of procedural appropriateness (Tyler, 2006; Bühlmann and Kunz, 2011).

Recent studies have identified factors influencing judicial AI acceptance across diverse contexts (Fine et al., 2025; Grimm et al., 2024). Research has examined individual differences—age, gender, cultural values, and technology attitudes—revealing meaningful variations (de la Osa and Remolina, 2024). Procedural justice constructs including fairness perceptions and system trust predict acceptance (Tyler, 2006; Tyler and Huo, 2002; Versteeg and Ginsburg, 2017).

However, existing research has focused on identifying influential factors while neglecting cognitive categorization processes in legal disputes. This represents a theoretical gap: citizens' AI acceptance depends not only on technology attitudes but also on how they cognitively organize disputes. How citizens intuitively organize legal domains before evaluating technological appropriateness remains unexplored.

## 1.2 Psychological dimensions of legal disputes: theoretical background

Legal disputes vary along psychological dimensions that may influence evaluation of decision-making processes. We propose that disputes can be characterized along dimensions reflecting interpersonal relationships versus institutional procedures. Social domain theory (Turiel, 1983) distinguishes moral transgressions involving harm between individuals from conventional violations of institutional rules, reflecting broader differences in mental representation (Nucci, 2001).

Construal level theory (Trope and Liberman, 2010) offers a complementary perspective: psychologically proximal situations evoke concrete, contextualized processing rich in emotional detail, whereas distant situations promote abstract, rule-based processing. Applied to legal contexts, interpersonal disputes may be construed concretely—as unique situations requiring empathetic understanding—whereas institutional disputes may be construed abstractly—as rule application amenable to algorithmic processing.

Categorization theory (Rosch, 1975; Murphy, 2004) that individuals organize complex domains into meaningful categories based on perceived similarity and functional relevance. These cognitive structures may guide evaluations of appropriate decision-making processes. Together, these frameworks suggest psychological features of dispute content may systematically influence acceptability judgments for AI versus human adjudication. This perspective builds on classic work on prototype structure and graded category membership (Rosch and Mervis, 1975; Rosch, 1978).

## 1.3 Integrating contextual and individual difference perspectives

We propose a framework wherein psychological characteristics of disputes (interpersonal involvement, emotional salience, institutional structure) are associated with acceptability judgments for AI versus human adjudicators. These associations may operate through cognitive categorization, emotional responses, moral intuitions, and procedural fairness perceptions (see Figure 1).

Individual differences—personality traits, gender, and AI-specific attitudes—may influence these associations through several mechanisms. They may shape how dispute features are perceived, moderate relationships between characteristics and judgments, or exert direct effects independent of content. This framework acknowledges that judgments reflect the interplay of situational and person characteristics, extending technology acceptance models by incorporating psychological features of decision contexts This aligns with frameworks emphasizing person-situation interactions in social judgment (Mischel and Shoda, 1995) and domain-specificity in moral cognition (Turiel, 1983). Dual-process perspectives further suggest that intuitive and deliberative processes may be differentially engaged depending on task characteristics and contextual framing (Evans, 2008; Evans and Stanovich, 2013).

## 1.4 Theoretical expectations regarding dispute dimensions

We expect acceptability may vary systematically with psychological characteristics of disputes. Disputes involving interpersonal relationships and emotional complexity may elicit stronger preferences for human adjudication, while disputes centered on institutional rules may show comparatively higher AI acceptance. This derives from the proposition that interpersonal disputes may be construed as requiring empathetic understanding, whereas institutional disputes may be viewed as amenable to rule-based consistency.

This dimensional distinction represents a theoretical prediction to be examined empirically. The Japanese legal context offers an appropriate setting, given its emphasis on consensus-building, procedural consistency, and hierarchical decision-making. Cultural emphasis on *wa* (harmony) may heighten sensitivity to distinctions between rule-based procedures and relationship management. Cross-cultural research will be necessary to assess universal versus culturally specific patterns.

The Japanese legal context offers a particularly appropriate setting for this investigation. Japan's legal system emphasizes consensus-building, procedural consistency, and hierarchical decision-making. The cultural emphasis on *wa* (harmony) and collective decision-making establishes distinctive frameworks for evaluating procedural appropriateness. These features may heighten sensitivity to distinctions between rule-based institutional procedures and interpersonal relationship management. Additionally, Japan's lay judge system (*saiban-in seido*) demonstrates public engagement with questions about appropriate decision-making processes in judicial contexts. Cross-cultural comparative research will be necessary to assess which findings reflect universal psychological structures versus culturally specific patterns of legal reasoning.

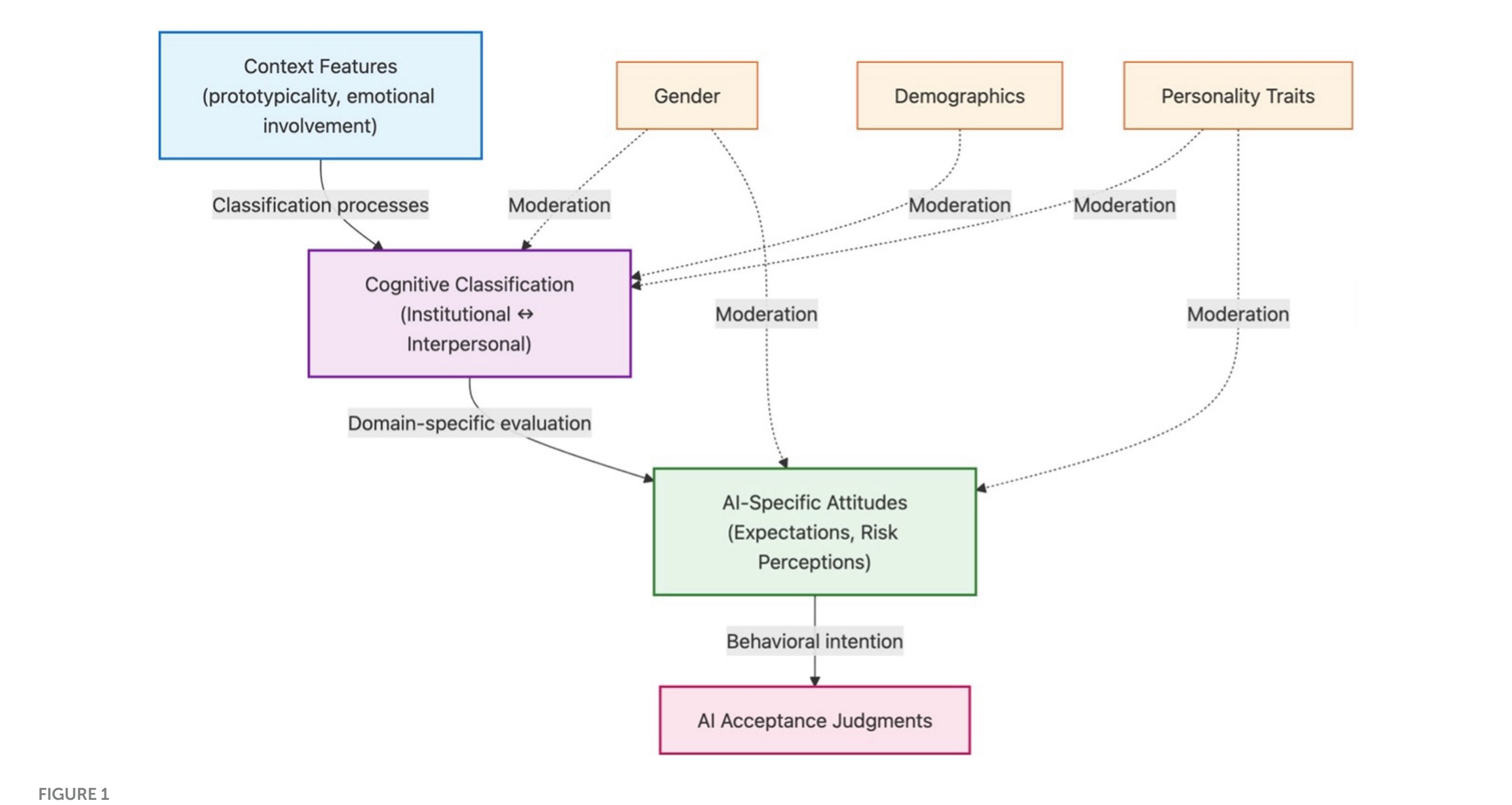

FIGURE 1
Conceptual framework integrating contextual and individual difference perspectives in AI acceptance. The framework illustrates associations between psychological characteristics of disputes and acceptability judgments for AI versus human adjudicators. Contextual features (prototypicality, emotional involvement) may influence how disputes are construed, which in turn may shape acceptability judgments. Individual differences influence acceptance through multiple pathways: moderating associations between dispute characteristics and acceptability, and exerting direct effects through domain-specific attitudes. Solid arrows indicate hypothesized associations tested in the present research; dashed arrows indicate potential moderating influences. These associations may operate through multiple psychological processes including categorization, emotional responses, moral intuitions, and procedural fairness perceptions. This model emphasizes that classification operates as an intermediate mechanism rather than a direct determinant of acceptance, and that individual differences influence acceptance through multiple pathways, including both classification-based and direct routes.

## 2 Research questions and hypotheses

Building on theoretical frameworks from the Introduction, this research examines whether acceptability judgments for AI versus human adjudication vary systematically with psychological characteristics of dispute content. Rather than treating acceptance as a unitary attitude toward algorithmic decision-making, we assess whether preference patterns correspond to theoretically meaningful dimensions—particularly the interpersonal-relational versus institutional-procedural distinction predicted by moral psychology and construal level theory.

This research examines whether acceptability for AI versus human adjudication varies systematically with psychological characteristics of dispute content. Rather than treating acceptance as unitary, we assess whether preferences correspond to theoretically meaningful dimensions—particularly interpersonal-relational versus institutional-procedural distinctions predicted by moral psychology and construal level theory.

### 2.1 Research questions

We organize the investigation around four research questions.

Research Question 1: What dimensional structure characterizes citizens' acceptability judgments across diverse legal disputes?

Research Question 2: Does this dimensional structure replicate across independent samples and exhibit stability across methodological approaches?

Research Question 3: Do experimentally manipulated contextual features—emotional involvement and prototypicality—systematically shift acceptability judgments?

Research Question 4: Do personality traits and AI-specific attitudes moderate contextual effects on acceptability judgments, or exert direct influences independent of dispute characteristics?

### 2.2 Hypotheses

#### 2.2.1 Hypothesis 1 (H1): two-dimensional structure

Acceptability judgments will exhibit a systematic two-dimensional structure distinguishing interpersonal-relational disputes (involving personal relationships and emotional dynamics) from institutional-procedural disputes (involving formal rules and organizational structures).

##### 2.2.1.1 Theoretical basis

Moral psychology distinguishes moral from conventional violations (Turiel, 1983). Construal level theory posits that proximal situations evoke concrete processing, while distal situations promote abstract processing (Trope and Liberman, 2010).

### 2.2.2 Hypothesis 2 (H2): dimensional variation in acceptability

Acceptability for AI adjudication will be higher for institutional-procedural disputes than for interpersonal-relational disputes. Disputes involving formal rules and organizational actors will elicit comparatively higher AI acceptance, reflecting perceived alignment between algorithmic processing and procedural consistency. Disputes involving personal relationships and emotional harm will elicit stronger preferences for human judgment, reflecting the perceived requirement for empathetic understanding.

#### 2.2.2.1 Theoretical basis

Perceived fit between technology characteristics and task requirements predicts acceptance (Davis, 1989; Venkatesh and Davis, 2000).

### 2.2.3 Hypothesis 3 (H3): contextual framing effects

Experimentally manipulated contextual features will systematically shift acceptability judgments. Disputes framed with high emotional involvement will elicit stronger preferences for human judgment. Disputes framed as prototypical (common, precedent-rich) will elicit higher AI acceptance than non-prototypical disputes.

#### 2.2.3.1 Theoretical basis

Construal level theory predicts emotional salience promotes concrete construal, increasing perceived need for human judgment (Trope and Liberman, 2010). Prototypical exemplars facilitate abstract processing (Rosch, 1975).

### 2.2.4 Hypothesis 4 (H4): individual difference moderators and direct effects

Personality traits and AI-specific attitudes will influence acceptability through multiple pathways: (a) moderating contextual effects, and (b) exerting direct effects independent of dispute characteristics.

### 2.2.5 H4a (personality traits)

Agreeableness will predict preference for human judgment, particularly in interpersonal disputes. Conscientiousness will predict acceptance of AI in procedurally structured disputes. Openness to experience will moderate emotional framing effects.

### 2.2.6 H4b (gender interactions)

Gender may moderate personality effects and contextual influences.

### 2.2.7 H4c (AI-specific attitudes as primary predictors)

Domain-specific attitudes—particularly expectations about AI capabilities and concerns about AI risks—will constitute the strongest proximal predictors of acceptability, mediating effects of personality traits.

#### 2.2.7.1 Theoretical basis

Big Five traits influence technology attitudes (Devaraj et al., 2008). The Theory of Planned Behavior positions domain-specific attitudes as proximal determinants, with personality as a distal influence (Ajzen, 1991).

## 2.3 Overview of studies

Study 1 assesses whether acceptability judgments exhibit systematic dimensional structure corresponding to theoretically meaningful features of dispute content. Using exploratory factor analysis on ratings across 46 legal dispute vignettes, we examine whether preference patterns align with the interpersonal-institutional distinction predicted by moral psychology and construal level theory.

Study 2 tests whether the dimensional structure observed in Study 1 replicates in an independent sample using confirmatory approaches, and whether experimentally manipulated contextual features—emotional involvement and prototypicality—systematically shift acceptability judgments. Study 2 also examines interactions between contextual features and individual difference variables (personality traits, AI-specific attitudes, gender).

These studies assess associations between dispute characteristics and acceptability judgments rather than directly measuring underlying psychological processes. While we interpret findings through categorization and construal level theories, alternative mechanisms—emotional responses, moral intuitions, and fairness perceptions—may contribute to observed patterns.

This two-study design follows established methodological principles for exploratory-confirmatory research sequences (Fabrigar et al., 1999). Study 1's exploratory approach is appropriate when latent structures remain unknown, using data-driven factor analysis to uncover dimensional organization without imposing predetermined structures. The resulting dimensional structure provides the empirical foundation for Study 2's hypothesis-driven investigation. Study 2's experimental design then tests whether this dimensional structure replicates in an independent sample, responds to theoretically motivated manipulations, and interacts with individual characteristics in predictable patterns. This sequential approach balances discovery and confirmation, allowing initial exploration while subjecting emergent patterns to rigorous experimental testing.

# 3 Study 1

## 3.1 Purpose

Study 1 examined whether acceptability judgments for AI versus human adjudication exhibit systematic dimensional structure across diverse legal disputes. Existing research indicates variation in acceptance across contexts (Fine et al., 2025; Sourdin, 2018), yet systematic examination of coherent psychological dimensions remains absent.

Drawing on moral psychology and construal level theory, we assessed whether preference patterns align with interpersonal-relational versus institutional-procedural distinctions. Participants evaluated 46 vignettes spanning diverse legal domains, enabling identification of systematic patterns rather than idiosyncratic responses.

## 3.2 Methods

### 3.2.1 Participants

Recruitment proceeded through Freeasy, an online survey platform with a nationally representative Japanese panel. Stratified quota sampling achieved demographic balance across gender and age (20–70). Following data-quality criteria, the final sample comprised 1,384 participants (733 women, 651 men; $M_{age} = 51.06$, SD = 15.67).

Sample size considerations. We targeted approximately 1,400 responses based on established factor analysis guidelines. Recommendations suggest minimum samples of 300 (Tabachnick and Fidell, 2012) and subject-to-item ratios of 10:1 (Costello and Osborne, 2005). The final sample (*N*/item ratio ≈ 30:1) provided adequate stability.

#### 3.2.1.1 Preregistration

We adopted an exploratory design reflecting nascent knowledge of how dispute characteristics shape judgments. Exploratory research suits unknown latent structures (Fabrigar et al., 1999), providing foundations for Study 2's confirmatory investigation.

### 3.2.2 Materials

#### 3.2.2.1 Vignette construction

Materials consisted of 46 single-sentence vignettes depicting diverse legal disputes. A social psychologist with legal training developed vignettes through a systematic multistage process, encompassing criminal and civil domains: family conflicts, traffic violations, theft, regulatory violations, employment disputes, and corporate misconduct. A second psychologist reviewed materials for clarity.

Single-sentence presentation minimizes processing demands while preserving case characteristics (Sweller et al., 1998). Scenarios reflected citizens' intuitive perceptions rather than formal jurisprudential distinctions.

### 3.2.3 Measurement of dependent variables

Participants evaluated each vignette using a four-point forced-choice scale: 1 = definitely AI, 2 = relatively AI, 3 = relatively human, 4 = definitely human. This approach eliminated neutral response options, compelling clear preferences. This format enhanced discrimination but eliminated the option to express equivalence. Study 2 employed a five-point scale including a midpoint to address this limitation. This measurement strategy is consistent with prior work on latent structure and construct representation in judgment research (Brown and Maydeu-Olivares, 2011), as well as studies on human interaction with automated systems (Parasuraman and Riley, 1997).

### 3.2.4 Procedure

Survey administration proceeded via Freeasy's secure platform following informed consent. Participants completed 48 items (46 vignettes plus two attention-checks) in random order. Randomization controlled for sequence effects. The interface supported desktop and mobile platforms with no time limits.

The survey included: (1) demographic information collection, (2) 46 vignettes in randomized order with acceptability ratings, (3) two attention-check items screening for inattentive responding.

We computed aggregated acceptability scores by summing raw responses within each factor (institutional vs. interpersonal) to obtain stable indices of acceptability tendencies. Because participants rated vignettes on 1–4 scales across numerous items per factor, composite values naturally exceeded the original scale, producing totals near or above 100. These aggregated indices represent cumulative classification strength across multiple scenarios. This aligns with conventional multi-item vignette research, where summing raw responses within theoretically derived clusters provides robust indicators of cognitive tendencies (DeVellis and Thorpe, 2022; Streiner, 2003).

### 3.2.5 Ethics

The Kansai University Graduate School of Psychology Ethics Review Committee approved all procedures in August 2023 (#338). Anonymity was ensured through randomly generated identifiers with encrypted data transmission and storage. The study was not preregistered, given its exploratory design, though our research plan and question items were presented in the ethical consideration form.

### 3.2.6 Rationale for exclusion criteria and data quality

Quality control integrated embedded attention checks and Freeasy's detection protocols. We excluded participants failing either attention-check item, resulting in 616 exclusions from 2,000 initial participants, yielding 1,384 respondents (31% exclusion rate). This falls within expected ranges for stringent online survey quality control (Miura and Kobayashi, 2015; Zhang et al., 2020), particularly when assessing careful reading of textual materials.

The 31% failure rate aligns with methodological baselines. Berinsky et al. (2014) reported 34–41% failure rates, Esch et al. (2025) documented 32.4%, and Alvarez et al. (2019) reported comparable rates, suggesting failure rates near 30% represent predictable features of unsupervised online data collection.

Attention check design substantially influences these rates. Anduiza and Galais (2017) demonstrated failure rates from 5–9% for simpler checks to approximately 60% for demanding items. Our moderate-difficulty checks yielded results consistent with the 30% threshold (Thomas and Clifford, 2017; Chandler et al., 2014). Following Berinsky et al. (2014), we distinguished respondents by attention level, balancing internal and external validity. The exclusion rate did not compromise statistical power, as evidenced by consistent effect patterns and robust factor structure.

## 3.3 Results

### 3.3.1 Descriptive statistics and classification patterns

Table 1 presents descriptive statistics for all 46 vignettes. Mean acceptance scores ranged from 2.46 (traffic accidents) to 3.06 (marital discord), indicating substantial variation in citizen preferences. Standard deviations ranged from 0.694 to 0.862, revealing meaningful differences in consensus levels. Disputes involving regulatory violations clustered toward lower means with moderate variability, indicating relatively consistent preferences for AI. Scenarios addressing family relationships, violent crimes, and child welfare consistently produced means exceeding 3.0, indicating broad consensus favoring human judgment.

### 3.3.2 Exploratory factor analysis

Exploratory factor analysis with promax rotation extracted two factors accounting for 61.6% of variance (Factor 1: 57.3%, Factor 2: 4.2% after rotation) (Table 2). Factor loadings demonstrated clear simple structure, with 42 of 46 items loading ≥0.40 on a single factor.

TABLE 1 Descriptive statistics for each question item.

| Item | Mean | SD |
|---|---|---|
| Q01 Valuables were stolen from my neighbor | 2.940 | 0.764 |
| Q02 Motorists ran a red light and crashed into each other | 2.460 | 0.851 |
| Q03 One got into a fight and hurt someone else | 3.010 | 0.700 |
| Q04 One bought a product on the internet but did not receive it | 2.540 | 0.829 |
| Q05 One was slandered on the internet | 2.700 | 0.833 |
| Q06 A married couple had trouble and they filed for divorce | 3.060 | 0.694 |
| Q07 A public official received a bribe in connection with their duties | 2.760 | 0.846 |
| Q08 One bought goods from a seller in person, but the quality of the goods was lower than described at the time of purchase | 2.770 | 0.787 |
| Q09 One trespassed on someone else's land without permission | 2.780 | 0.790 |
| Q11 One drank alcohol and caused a commotion in the park | 2.900 | 0.749 |
| Q12 Someone was violent toward their child | 3.030 | 0.769 |
| Q13 One interfered with the progress of a trial in court | 2.780 | 0.793 |
| Q14 A group of drug traffickers was caught | 2.750 | 0.839 |
| Q15 One gave false testimony at a trial | 2.710 | 0.838 |
| Q16 Prostitution was conducted in an illegal brothel | 2.820 | 0.795 |
| Q17 A suspect was arrested for murdering another person | 3.020 | 0.767 |
| Q18 A person profited from copying another person's copyrighted work without permission | 2.510 | 0.842 |
| Q19 The advertisement of a product is different from the actual product | 2.560 | 0.817 |
| Q20 One got followed or harassed by a stalker | 2.920 | 0.801 |
| Q21 Someone forced somebody to repay a debt through illegal means | 2.800 | 0.811 |
| Q22 A worker's health was damaged due to overwork | 2.840 | 0.826 |
| Q23 One obstructed a police officer in the performance of their duties | 2.790 | 0.785 |
| Q24 One struck a person in a fight | 2.940 | 0.727 |
| Q25 One helped a person who was asked to commit suicide | 2.990 | 0.729 |
| Q26 One improperly used another company's patent or trademark | 2.520 | 0.832 |
| Q27 One drove a car after consuming alcohol | 2.710 | 0.858 |
| Q28 An underage person consumed alcohol | 2.760 | 0.810 |
| Q29 A company failed to comply with environmental laws and polluted the air or water | 2.610 | 0.843 |
| Q30 A company leaked the personal information of a customer | 2.540 | 0.862 |
| Q31 A company discriminated against women in terms of employment and wages | 2.740 | 0.817 |
| Q32 A municipality made a decision that violated an individual's human rights | 2.750 | 0.810 |
| Q33 A newly built house was defective and I sought repairs or compensation | 2.690 | 0.847 |
| Q35 My supervisor or co-worker made sexually explicit comments or actions | 2.800 | 0.822 |
| Q36 A building collapsed and people were injured or killed | 2.800 | 0.823 |
| Q37 An employer fired a person for illegal reasons | 2.800 | 0.797 |
| Q38 One was involved in an online scam | 2.480 | 0.852 |
| Q39 A violent incident occurred in a public place and there were victims | 2.880 | 0.751 |
| Q40 One failed to fulfill its contractual obligations | 2.640 | 0.820 |
| Q41 A restaurant was caught violating food safety laws | 2.700 | 0.809 |
| Q42 A fraudster provided false information to obtain a loan | 2.670 | 0.836 |
| Q43 A dispute over custody of children arose between a divorcing couple | 3.000 | 0.750 |
| Q44 A company was alleged to have been involved with antisocial forces | 2.740 | 0.799 |
| Q45 A robbery was committed with a weapon | 2.800 | 0.841 |
| Q46 The accused sexually assaulted a victim | 2.890 | 0.829 |

*(Continued)*

TABLE 1 (Continued)

| Item | Mean | SD |
|---|---|---|
| Q47 Someone trafficked in large quantities of narcotics | 2.710 | 0.854 |
| Q48 A group committed a major fraud | 2.670 | 0.861 |

The table presents the mean values and standard deviations (SD) for each of the 48 question items used in the survey. Items cover a wide range of interpersonal, legal, and social transgressions (e.g., theft, bribery, violence, fraud, discrimination, and corporate misconduct). These statistics provide an overview of the relative frequency or perceived severity of each scenario as evaluated by participants. $N = 1{,}384$. Items Q10 and Q34 were included as attention check questions and are not reported here.

The dimensional structure aligned with theoretical distinctions from moral psychology and construal level theory. Factor 1 items involved rule-based institutional contexts and procedural disputes (e.g., patent infringement, regulatory violations). Factor 2 items centered on interpersonal relationships and emotionally salient situations (e.g., child custody, divorce, violent crimes). We label these "Institutional-Procedural Disputes" (Factor 1) and "Interpersonal-Relational Disputes" (Factor 2).

Preliminary checks confirmed suitability: KMO = 0.99; Bartlett's test, $\chi^2(1035) = 57419.81$, $p < 0.001$. Parallel analysis and scree plot converged on a two-factor solution. Observed eigenvalues (26.36, 1.96) substantially exceeded random thresholds.

Factor 1 encompassed disputes with formal regulatory frameworks and institutional actors. Factor 2 encompassed direct interpersonal harm and family conflicts. The two factors correlated moderately ($r = 0.805$), indicating distinct yet integrated constructs. Some scenarios showed meaningful cross-loadings (e.g., armed robbery: $\lambda_1 = 0.654$, $\lambda_2 = 0.587$), suggesting sensitivity to multifaceted dispute characteristics.

### 3.3.3 Quadrant analysis

We conducted quadrant analysis employing median splits (Table 3) to examine the joint distribution of mean scores (indicating preference direction: AI- vs. human-oriented) and standard deviations (indicating consensus level vs. disagreement). The median mean score (2.765) distinguished between scenarios generating AI-oriented and human-oriented classifications. The median standard deviation (0.817) differentiated between scenarios characterized by high classification consensus and substantial disagreement among respondents.

The institutional-consensus quadrant encompassed scenarios that generated AI-oriented classifications with broad consensus. These included employment discrimination (M = 2.63, SD = 0.81, $d = 0.28$ below the median) and violations of food safety regulations (M = 2.49, SD = 0.72, $d = 0.32$ below the median). These scenarios activated classification schemes emphasizing procedural consistency and objective evaluation.

Conversely, the interpersonal-consensus quadrant included scenarios producing human-oriented classification with broad consensus. These encompassed disputes regarding child custody (M = 2.97, SD = 0.74, $d = 0.20$ above median) and cases of child abuse (M = 2.98, SD = 0.73, $d = 0.21$ above median). These scenarios activated classification schemes emphasizing empathic understanding and relational sensitivity.

Boundary cases occupied the disagreement quadrants. The AI-disagreement quadrant contained scenarios exhibiting AI acceptance but generating substantial response variability. Traffic accidents exemplified this pattern (M = 2.46, SD = 0.86, indicating 0.31 below median mean but elevated disagreement, $d = 0.38$). The human-disagreement quadrant included scenarios in which participants favored human adjudication but with substantial response variability. Harm due to workplace overwork illustrated this pattern (M = 2.94, SD = 0.84, indicating 0.17 above median mean with elevated disagreement, $d = 0.21$).

These boundary cases suggest that certain dispute types activate competing classification schemes or generate genuine disagreement regarding appropriate procedural allocation. This variability potentially reflects heterogeneity in citizens' implicit theories about dispute types.

### 3.3.4 Demographic predictors

Multiple regression analyses examined demographic influences. For institutional disputes, the model explained 1.2% of variance, $F(3, 1{,}380) = 5.63$, $p < 0.001$. Age emerged as a significant negative predictor ($\beta = -0.098$, $p < 0.001$, $\eta^2 = 0.010$), indicating older participants displayed greater AI acceptance. For interpersonal disputes, the model explained 0.7% of variance, $F(3, 1{,}380) = 3.37$, $p = 0.018$, with education as the significant predictor ($\beta = 0.064$, $p = 0.019$, $\eta^2 = 0.004$).

### 3.3.5 Aggregate score analysis

We computed composite scores as mean ratings for each factor. Aggregate scores revealed striking differentiation: participants expressed significantly stronger preferences for human judgment in interpersonal disputes (M = 2.89, SD = 0.61) compared with institutional disputes (M = 2.65, SD = 0.65), $t(1383) = -31.27$, $p < 0.001$, $d = -0.84$, 95% CI (−0.902, −0.779). Although both means exceed the scale midpoint (2.5), indicating overall preference for human judgment, the effect size reflects pronounced within-participant consistency in dispute-type differentiation.

These results reveal that citizens demonstrate differential but consistent human judgment preferences across dispute types rather than categorical AI acceptance. Even in institutional disputes, participants retained a slight preference for human involvement. This suggests that cognitive classification operates to modulate human judgment preferences rather than create a binary choice favoring AI in specific domains.

### 3.3.6 Robustness of results across attention-check thresholds

To evaluate the sensitivity of the factor-analytic findings to variations in data-quality thresholds, we conducted supplementary analyses employing progressively relaxed exclusion criteria. We examined three quality-screening scenarios: (a) the strict criterion, requiring correct responses to both attention-check items; (b) a lenient criterion, requiring correct responses to at least one attention-check item; and (c) an unscreened dataset, retaining all respondents regardless of attention-check performance. This hierarchical approach permitted assessment of both the structural robustness of the two-factor solution and the sensitivity of effect estimates to quality-control decisions.

TABLE 2 The results of the exploratory factor analysis on AI-human scales.

| Item | Factor 1 | Factor 2 |
|---|---|---|
| Q30 A company leaked the personal information of a customer | 0.976 | |
| Q26 One improperly used another company's patent or trademark | 0.952 | |
| Q38 One was involved in an online scam | 0.933 | |
| Q18 A person profited from copying another person's copyrighted work without permission | 0.922 | |
| Q19 The advertisement of a product is different from the actual product | 0.900 | |
| Q29 A company failed to comply with environmental laws and polluted the air or water | 0.791 | |
| Q42 A fraudster provided false information to obtain a loan | 0.708 | |
| Q40 One failed to fulfill its contractual obligations | 0.704 | |
| Q02 Motorists ran a red light and crashed into each other | 0.623 | |
| Q47 Someone trafficked in large quantities of narcotics | 0.607 | |
| Q33 A newly built house was defective and I sought repairs or compensation | 0.597 | |
| Q04 One bought a product on the internet but did not receive it | 0.595 | |
| Q48 A group committed a major fraud | 0.593 | |
| Q41 A restaurant was caught violating food safety laws | 0.576 | |
| Q27 One drove a car after consuming alcohol | 0.553 | |
| Q32 A municipality made a decision that violated an individual's human rights | 0.542 | |
| Q28 An underage person consumed alcohol | 0.542 | |
| Q44 A company was alleged to have been involved with antisocial forces | 0.517 | |
| Q31 A company discriminated against women in terms of employment and wages | 0.515 | |
| Q14 A group of drug traffickers was caught | 0.471 | |
| Q15 One gave false testimony at a trial | 0.468 | |
| Q13 One interfered with the progress of a trial in court | 0.445 | |
| Q05 One was slandered on the internet | 0.439 | |
| Q08 One bought goods from a seller in person, but the quality of the goods was lower than described at the time of purchase | 0.417 | |
| Q45 A robbery was committed with a weapon | 0.390 | 0.441 |
| Q12 Someone was violent toward their child | | 0.911 |
| Q06 A married couple had trouble and they filed for divorce | | 0.901 |
| Q17 A suspect was arrested for murdering another person | | 0.844 |
| Q03 One got into a fight and hurt someone else | | 0.809 |
| Q43 A dispute over custody of children arose between a divorcing couple | | 0.808 |
| Q24 One struck a person in a fight | | 0.766 |
| Q20 One got followed or harassed by a stalker | | 0.733 |
| Q25 One helped a person who was asked to commit suicide | | 0.726 |
| Q46 The accused sexually assaulted a victim | | 0.716 |
| Q39 A violent incident occurred in a public place and there were victims | | 0.688 |
| Q01 Valuables were stolen from my neighbor | | 0.579 |
| Q11 One drank alcohol and caused a commotion in the park | | 0.570 |
| Q07 A public official received a bribe in connection with their duties | | 0.502 |
| Q37 An employer fired a person for illegal reasons | | 0.488 |
| Q23 One obstructed a police officer in the performance of their duties | | 0.484 |
| Q16 Prostitution was conducted in an illegal brothel | | 0.484 |
| Q21 Someone forced somebody to repay a debt through illegal means | | 0.469 |
| Q22 A worker's health was damaged due to overwork | | 0.461 |
| Q36 A building collapsed and people were injured or killed | | 0.455 |

*(Continued)*

TABLE 2 (Continued)

| Item | Factor 1 | Factor 2 |
|---|---|---|
| Q35 My supervisor or co-worker made sexually explicit comments or actions | | 0.416 |
| Q09 One trespassed on someone else's land without permission | | 0.381 |

This table shows the factor loadings of each item on two extracted factors obtained through exploratory factor analysis ($N = 1{,}384$). Items associated with corporate or societal misconduct (e.g., data leakage, fraud, environmental violations, or contractual breaches) loaded primarily on Factor 1, whereas items related to interpersonal and violent behaviors (e.g., assault, homicide, sexual misconduct, or child abuse) loaded on Factor 2. Loadings greater than 0.40 are reported.

Factor structures remained highly consistent across all three quality-threshold conditions. Correlations among factor loadings across quality conditions ranged from 0.96 to 0.99 for the institutional factor and from 0.97 to 0.99 for the interpersonal factor, values substantially exceeding the conventional 0.90 threshold for confirming factorial identity. These findings indicate that the underlying two-factor cognitive structure proves empirically robust and recoverable even under substantially relaxed quality-screening conditions.

Despite this structural consistency, measurement precision and effect estimates differed markedly across quality thresholds. The strict-quality sample ($N = 1{,}384$) demonstrated strong internal consistency for both factors (institutional: $\alpha = 0.972$; interpersonal: $\alpha = 0.967$) and yielded a medium-to-large within-participant effect, $t(1383) = -31.27$, $p < 0.001$, $d = -0.84$, 95% CI (−0.902, −0.779). The lenient-criterion sample ($N = 1{,}730$) exhibited comparable internal consistency (institutional: $\alpha = 0.977$; interpersonal: $\alpha = 0.973$) yet produced a notably smaller effect, $t(1729) = -31.37$, $p < 0.001$, $d = -0.754$, 95% CI (−0.807, −0.701), representing a 10.2% reduction in effect magnitude relative to the strict-quality condition. The unscreened sample ($N = 2{,}000$) demonstrated similar internal consistency (institutional: $\alpha = 0.976$; interpersonal: $\alpha = 0.971$) but yielded the most attenuated effect, $t(1999) = -30.84$, $p < 0.001$, $d = -0.694$, 95% CI (−0.738, −0.641), representing a combined 17.4% reduction from the strict-quality condition. This progressive attenuation of the theoretically expected effect with declining data quality indicates that inattentive respondents systematically inflated error variance and obscured meaningful differentiation across dispute types.

Notably, the internal consistency estimates remained surprisingly robust across all quality conditions, with $\alpha$ values ranging from 0.967 to 0.977. This pattern suggests that the factor structures themselves maintain integrity even with the inclusion of low-engagement respondents. However, the marked decline in effect size precision—from $d = -0.84$ to $d = -0.69$—despite maintained factor structure demonstrates that inattentive respondents introduce systematic noise into participant responses. This noise attenuates behavioral differentiation between dispute types without destabilizing the underlying cognitive classification structure.

Collectively, these sensitivity analyses demonstrate that although the two-factor structure proves statistically recoverable and maintains strong internal consistency across quality-control conditions, the precision and magnitude of substantive effects depend critically on excluding low-engagement respondents. The strict attention-check criterion, therefore, represents not merely a conservative procedural choice but a necessary prerequisite for obtaining effect estimates that accurately reflect cognitive classification processes. The consistency of the factor structure, combined with the substantial improvement in effect-size precision under strict quality criteria, strongly supports the robustness and validity of the conclusions we report in this study.

## 3.4 Discussion

### 3.4.1 Classification patterns, consensus, and boundary effects

This two-factor structure should be interpreted as a descriptive organization of acceptability judgments that systematically covaries with evaluations of AI suitability across dispute types. Consistent modulation by contextual manipulations in Study 2 is compatible with a classification-based account, but does not by itself establish classification as the underlying psychological mechanism. Study 1 revealed systematic variation in acceptability judgments corresponding to theoretically meaningful features of dispute content. The two-dimensional structure distinguishing interpersonal-relational from institutional-procedural disputes demonstrated stability and interpretability. Citizens exhibited stronger preferences for human judgment in interpersonal disputes (M = 2.89) compared to institutional disputes (M = 2.65), with substantial within-participant differentiation ($d = -0.84$).

The quadrant analysis revealed meaningful patterns of consensus and disagreement. Institutional disputes, such as regulatory violations and employment discrimination, clustered in the AI-consensus quadrant, while interpersonal disputes, such as child custody and child abuse, clustered in the human-consensus quadrant. Boundary cases occupying disagreement quadrants—such as traffic accidents and workplace overwork cases—suggest that certain dispute types may activate competing classification schemes or generate genuine uncertainty regarding appropriate procedural allocation.

These patterns provide initial evidence that psychological characteristics of dispute content systematically influence acceptability judgments. However, the dimensional structure emerged from exploratory analysis and represents descriptive organization of response patterns. Whether this structure reflects cognitive classification processes, emotional responses, moral intuitions, or other psychological mechanisms requires further investigation through experimental manipulation and direct process measurement, which Study 2 addresses.

### 3.4.2 Interpretation and scope

The observed dimensional structure provides descriptive evidence that acceptability judgments vary systematically with dispute characteristics. However, Study 1's correlational design does not establish whether classification processes causally influence acceptance judgments, whether both are driven by common underlying factors (e.g., emotional responses or moral intuitions), or whether alternative causal sequences operate. This interpretation is consistent with research on procedural justice, which emphasizes the importance of perceived fairness and legitimacy in the acceptance of decision-making authorities (Lind and Tyler, 1988; Tyler, 1988). The robust dimensional structure establishes that citizens differentiate between dispute types in their acceptance judgments, but identifying the psychological mechanisms producing this differentiation requires experimental manipulation and direct process measurement, which Study 2 addresses.

TABLE 3 Items placed in four quadrants according to their mean scores and standard deviations.

| AI ← | Median → Human | |
|---|---|---|
| Q02 Motorists ran a red light and crashed into each other<br>Q04 I bought a product on the internet but did not receive it<br>Q05 I was slandered on the internet<br>Q07 A public official received a bribe in connection with his/her duties<br>Q14 A group of drug traffickers was caught<br>Q15 I gave false testimony at a trial<br>Q18 A person profited from copying another person's work without permission<br>Q26 You improperly used another company's patent or trademark<br>Q27 You drove a car after consuming alcohol<br>Q29 The company failed to comply with environmental laws and polluted the air or water<br>Q30 The company leaked the personal information of a customer<br>Q33 A newly built house was defective and I sought repairs or compensation<br>Q38 I was involved in an online scam<br>Q40 The company failed to fulfill its contractual obligations<br>Q42 A fraudster provided false information to obtain a loan<br>Q47 The group trafficked in large quantities of narcotics<br>Q48 The group committed a major fraud | Q22 A worker's health was damaged due to overwork<br>Q35 My supervisor or co-worker made sexually explicit comments or actions<br>Q36 A building collapsed and people were injured or killed<br>Q45 A robbery was committed with a weapon<br>Q46 The accused sexually assaulted the victim | ↑<br>Disagreement<br>SD |
| Q19 The advertisement of a product is different from the actual product<br>Q28 An underage person consumed alcohol<br>Q31 The company discriminated against women in terms of employment and wages<br>Q32 A municipality made a decision that violated an individual's human rights<br>Q41 A restaurant was caught violating food safety laws<br>Q44 A company was alleged to have been involved with antisocial forces | Q01 My valuables were stolen from my neighbor<br>Q03 I got into a fight and hurt someone else<br>Q06 Marital discord arose and the couple filed for divorce<br>Q08 I bought goods from a face-to-face seller, but the quality of the goods was lower than explained<br>Q09 I trespassed on someone else's land without permission<br>Q11 I drank alcohol and caused a commotion in the park<br>Q12 I was violent toward my child<br>Q13 I interfered with the progress of a trial in court<br>Q16 Prostitution was conducted in an illegal brothel<br>Q17 A suspect was arrested for murdering another person<br>Q20 I was followed or harassed by a stalker<br>Q21 I was forced to repay a debt through illegal means<br>Q23 I obstructed a police officer in the performance of his/her duties<br>Q24 I struck a person in a fight<br>Q25 I helped a person who was asked to commit suicide<br>Q37 My employer fired me for illegal reasons<br>Q39 A violent incident occurred in a public place and there were victims<br>Q43 A dispute over custody of children arose between a divorcing couple | Median<br>Agreement<br>↓ |

### 3.4.3 Evaluation of hypotheses and interpretation

Study 1 provides initial evidence regarding the latent structure of citizens' cognitive classifications and their relationship to judicial AI acceptance. Hypothesis 1 (H1) predicted that institutional and interpersonal disputes would emerge as two distinct, psychologically meaningful dimensions. The exploratory factor analyses strongly supported this prediction: the two-factor solution accounted for 61.6% of total variance, with institutional factor loadings clustering around formal regulatory breaches and contractual violations ($\lambda > 0.75$), and interpersonal loadings clustering around family relationships and violent crime ($\lambda > 0.72$). This meaningful semantic clustering indicates that citizens organize disputes according to fundamental characteristics—whether resolution requires procedural rule application or contextual understanding—providing evidence for coherent cognitive classification schemes.

Hypothesis 2 (H2) proposed that cognitive classification would predict AI acceptance. Study 1 supported this prediction: institutional disputes showed higher AI acceptance ($M = 2.65$, $SD = 0.65$) compared with interpersonal disputes ($M = 2.89$, $SD = 0.61$), with substantial differentiation [$d = -0.84$, 95% CI ($-0.902$, $-0.779$)]. Notably, citizens retained overall human-judgment preference even in institutional disputes, indicating that classification operates to modulate the strength of human preference rather than to create categorical AI acceptance. This pattern—consistent across 46 diverse vignettes and 1,384 participants—provides evidence that citizens' acceptance judgments reflect their cognitive organization of disputes and implicit theories about procedural requirements.

These findings support H1 and H2 and establish the two-factor classification structure as a robust foundation for further investigation. The dimensional structure demonstrates that citizens organize legal disputes along psychologically meaningful dimensions when evaluating AI versus human adjudication. However, the exploratory and correlational nature of Study 1 leaves several critical questions unanswered.

First, whether the dimensional structure replicates across independent samples and different measurement formats remains to be established. Second, whether contextual features theorized to influence cognitive construal—such as emotional involvement and prototypicality—systematically modulate acceptability judgments requires experimental investigation. Third, how individual differences in personality traits and AI-specific attitudes interact with dispute characteristics to shape acceptance patterns requires explicit examination. Study 2 addresses these questions through experimental manipulation and confirmatory analysis.

### 3.4.4 Implications for Study 2

Study 1 revealed systematic variation in acceptability judgments corresponding to theoretically meaningful features of dispute content. The two-dimensional structure distinguishing interpersonal-relational from institutional-procedural disputes demonstrated stability and interpretability. However, several limitations constrain interpretation.

First, the dimensional structure emerged from exploratory analysis of the same acceptability ratings it describes. While factor analysis is appropriate for uncovering latent structure, the resulting dimensions represent descriptive summaries of response patterns rather than direct evidence for underlying psychological processes. Second, the correlational design precludes causal inference regarding whether contextual features actively shape judgments or merely correlate with stable preferences. Third, the forced-choice response format enhanced discrimination but eliminated the option to express equivalence between AI and human adjudicators.

Study 2 addresses these limitations through experimental manipulation of contextual features and replication of the dimensional structure in an independent sample with a midpoint-inclusive response scale. The stable two-factor structure identified in Study 1 provides theoretical guidance for selecting contextual features to manipulate: prototypicality and emotional involvement represent dimensions theoretically predicted to influence how disputes are mentally represented. Boundary cases exhibiting intermediate acceptance and elevated variance offer particularly promising contexts for examining how contextual framing and individual differences modulate judgments.

## 4 Study 2: contextual features and individual differences in AI acceptance

Study 1 documented systematic variation in acceptability judgments across dispute types. Study 2 employs experimental manipulation to examine whether contextual features shift acceptability judgments and whether individual differences moderate these effects.

Three questions guide Study 2. First, does the dimensional structure replicate in an independent sample with a midpoint-inclusive response scale? Second, do experimentally manipulated contextual features—emotional involvement and prototypicality—influence acceptability judgments? Third, do individual differences in personality traits, AI-specific attitudes, and gender moderate contextual effects or exert direct influences?

This approach assesses whether situational features contribute to acceptance beyond stable individual differences. Study 2 examines associations between manipulated contextual features and acceptability judgments rather than directly testing cognitive mechanisms.

### 4.1 Purpose and rationale

Study 1 identified a robust two-dimensional structure organizing acceptability judgments for AI versus human adjudication, distinguishing interpersonal-relational from institutional-procedural disputes. This aligned with theoretical predictions from moral psychology (Turiel, 1983) and construal level theory (Trope and Liberman, 2010), demonstrating substantial within-participant differentiation ($d = -0.84$). However, Study 1's exploratory design left questions unresolved regarding psychological mechanisms.

Study 2 addresses three objectives, building on Study 1. First, we established replicability through confirmatory analysis in an independent sample. To assess whether dimensional structure depends on forced-choice responding, Study 2 employed a five-point scale with a neutral midpoint, allowing participants to express equivalence when preferences are ambiguous.

Second, we tested whether experimentally manipulated contextual features modulate acceptability judgments in predicted directions. If dimensional structure reflects cognitive classification processes, then contextual features influencing mental representation should shift acceptability patterns. We manipulated emotional involvement and prototypicality.

Emotional involvement was predicted to increase human judgment preference. Construal level theory posits that emotionally salient situations evoke concrete, contextualized processing (Trope and Liberman, 2010), whereas neutral situations promote abstract, rule-based processing. Emotional framing should activate construals emphasizing human empathy, reducing AI acceptance.

Prototypicality was predicted to moderate acceptance. Prototypical exemplars facilitate abstract, rule-based processing and increase algorithmic confidence (Rosch, 1975; Murphy, 2004). Non-prototypical cases should activate concrete construals emphasizing case-specific features, reducing AI acceptance.

Third, we examined whether contextual effects interact with individual differences: (a) Big Five personality traits, particularly agreeableness and conscientiousness; (b) AI-specific attitudes, including expectations and risk concerns; and (c) gender, which may moderate emotional framing responses (Venkatesh and Morris, 2000).

These objectives test whether Study 1's dimensional structure reflects stable psychological organization responding to contextual manipulation in theoretically meaningful ways.

## 4.2 Methods

### 4.2.1 Participants and sampling

Participants were recruited through Freeasy using stratified quota methodology employed in Study 1. Recruitment occurred in December 2024. Initial recruitment of 1,818 panelists incorporated oversampling to accommodate anticipated exclusions while ensuring adequate statistical power.

Application of exclusion criteria yielded a final sample of 596 participants (247 men, 349 women, $M_{age}$ = 55.45, SD = 17.56). The 67.2% exclusion rate exceeded that of Study 1 (31%) due to more demanding quality-control procedures necessitated by experimental design complexity.

#### 4.2.1.1 Sample size and power analysis

*A priori* power analysis using G*Power 3.1.9.6 (Faul et al., 2007) estimated the required sample size. We conservatively estimated the three-way interaction effect size at $f = 0.17$ ($\eta^2 \approx 0.028$). Under parameters ($\alpha = 0.05$, power = 0.80, $2 \times 2 \times 2$ design), G*Power indicated minimum required $N = 452$. Given the anticipated 25% additional exclusions, the recruitment target was 1,818.

#### 4.2.1.2 Preregistration status

This study was not preregistered. However, experimental design, manipulations, and primary hypotheses regarding three-way interactions were specified in advance based on Study 1. The analytical approach (ANOVA, PROCESS) was predetermined at the time of ethical consideration. Future research should employ formal preregistration to enhance transparency.

### 4.2.2 Quality control and exclusion criteria

Study 1 incorporated attention checks assessing careful reading, resulting in 30% exclusion. Study 2 maintained these checks while adding comprehension assessments targeting manipulation understanding. After experimental instructions, respondents answered two binary comprehension questions: (1) "Are the case types common occurrences or not?" and (2) "Do descriptions involve emotional elements or not?" Participants providing correct responses to both were retained; others were excluded.

This two-tiered approach addressed distinct invalid data sources. Standard attention checks identify respondents failing to read carefully—a baseline requirement for online surveys. Manipulation comprehension questions assess accurate processing of experimental framing, required for causal investigation. Attention check failures indicate insufficient reading engagement; comprehension failures indicate inadequate manipulation processing.

Although the 67.2% exclusion rate was substantial, it eliminated participants who were unable to engage with the experimental paradigm. Participants who could not answer straightforward comprehension questions would not reliably experience intended treatments, rendering data invalid for causal hypotheses. The strategy prioritized validity over sample size, which is appropriate for establishing causal mechanisms. Sensitivity analyses employing relaxed criteria yielded qualitatively equivalent factor structures and effects, supporting robustness and stringent standards.

The theoretical implication: causal inference requires participants accurately perceive intended manipulations. The practical implication: judicial AI communication must account for comprehension variation.

High exclusion rates suggest possible systematic selection effects. Excluded participants may differ from retained participants in digital literacy, survey familiarity, or technology attitudes. We compared demographic characteristics (age, gender, and education) between groups; distributions were broadly similar with minor differences in age and education, suggesting limited demographic distortion. However, unobserved factors—particularly technological familiarity and cognitive engagement—may differentiate the analytic sample from the general population, warranting cautious generalization, especially for populations with lower digital literacy or AI exposure. Future research should explore alternative quality-control strategies balancing internal validity with representativeness, such as adaptive comprehension training or tiered task complexity.

### 4.2.3 Experimental design

The study employed a $2 \times 2$ between-subjects factorial design examining two theoretically derived factors. The prototypicality factor contrasted prototypical legal disputes (common, precedent-rich cases) against non-prototypical disputes requiring case-specific judgment. The emotional involvement factor distinguished emotionally involving disputes emphasizing interpersonal emotions from emotion-neutral disputes presenting fact-based disagreements.

Random assignment employed stratified randomization, ensuring balanced demographic representation. Condition assignments: prototypical/emotionally involving ($N = 206$), prototypical/emotion-neutral ($N = 124$), non-prototypical/emotionally involving ($N = 141$), and non-prototypical/emotion-neutral ($N = 125$).

Participants received condition-specific instructions. The prototypical condition described disputes as "common legal matters with established precedents and standardized procedures," whereas the non-prototypical condition presented disputes as "rare legal matters requiring case-specific judgment." The emotionally involving condition emphasized "significant interpersonal emotions and personal suffering," whereas the emotion-neutral condition underscored "calm, fact-based disagreements requiring objective resolutions."

### 4.2.4 Materials and instrumentation

#### 4.2.4.1 Dispute vignettes and response scales

Experimental stimuli consisted of the same 46 vignettes employed in Study 1, ensuring direct comparability while enabling replication of cognitive classification structure. However, the response format was refined from Study 1's four-point forced-choice scale to a five-point Likert-type scale to accommodate experimental manipulation detection.

**4.2.4.1.1 Methodological rationale for scale format change.** The shift from four-point to five-point format reflects methodological priorities tailored to each study's objectives. Study 1 employed forced-choice (1 = definitely AI, 2 = relatively AI, 3 = relatively human, 4 = definitely human) to address Japanese respondents' tendency toward middle-category selection (Chen et al., 1995; Harzing, 2006; Si and Cullen, 1998). This maximizes clarity in identifying two-factor structure by compelling directional preferences, revealing latent cognitive dimensions.

We examined patterns of midpoint (neutral) responding following the scale modification. Across all items, mean midpoint selection was 27.2% (median = 20.0%), with a minority of participants exhibiting high usage (>50%). Critically, midpoint selection rates did not vary by emotional priming or typicality condition, nor by their interaction (all *p*-values >0.49), indicating that neutral responding was not systematically influenced by experimental manipulations. Sensitivity analyses excluding high-midpoint-use participants (>50%) produced substantively identical results for all main effects, confirming that findings are robust to potential neutral-response tendencies.

Study 2 sought to detect subtle contextual manipulation effects. Research indicates five-point or seven-point scales provide superior sensitivity for detecting moderate framing effects compared to forced-choice formats (Kühberger, 1998; Krosnick and Presser, 2010; Wetzel et al., 2020). The rationale: experimental manipulations creating genuine uncertainty about adjudicator suitability may manifest not only as shifts between opposing poles but also as reduced directional commitment.

The five-point format (1 = definitely AI, 2 = relatively AI, 3 = neither, 4 = relatively human, 5 = definitely human) introduces a "neither" response option, allowing participants whose preferences remain ambiguous to express ambivalence rather than forcing artificial choices. This neutral midpoint enables detection of manipulation-induced classification shifts that might manifest as reduced directional preference or shifts between poles. For example, if high emotional involvement creates uncertainty about whether algorithmic consistency or human empathy is more appropriate, participants might select "neither" rather than being forced to choose. Capturing genuine ambivalence is essential for understanding how contextual factors modulate classification processes.

This scale modification increases measurement sensitivity for detecting moderate contextual effects without eliminating forced-choice function. The five-point scale retains clear directional options while adding flexibility for genuine ambivalence, prioritizing ecological validity while maintaining constraint to prevent neutral-response bias.

The theoretical implication: cognitive classification operates as a robust psychological mechanism across measurement contexts, not merely as forced-choice artifact. The practical implication: assessing judicial AI acceptance may benefit from response formats permitting genuine ambivalence expression, particularly when AI suitability is contested.

#### 4.2.4.2 Assessment of individual differences

Personality traits were measured using the Japanese version of the Ten-Item Personality Inventory (TIPI-J; Oshio et al., 2012), providing concise Big Five assessment within online survey constraints. AI-specific attitudes were assessed using the Attitudes Toward AI Scale (Ota, 2023), comprising two 8-item subscales measuring AI expectations and risk perceptions on five-point scales. This instrument measures domain-specific evaluative orientations toward judicial AI rather than generic technology attitudes, enabling examination of domain-specific mediation pathways.

### 4.2.5 Procedure and analysis

After informed consent, participants completed three sequential phases: individual differences assessment, experimental manipulation and comprehension checking, and dispute evaluation across 46 vignettes. The survey required approximately 20–25 min, with interfaces optimized for desktop and mobile devices. Manipulation effectiveness was assessed through binary comprehension questions; correct responses on both were required for inclusion.

#### 4.2.5.1 Analytical strategy

The study employed univariate ANOVA examining experimental condition effects on AI acceptance, with prototypicality, emotional involvement, and gender as between-subjects factors. Exploratory factor analysis using principal axis factoring with promax rotation examined two-factor classification structure replicability (Fabrigar et al., 1999). Mixed-design ANCOVA examined dispute type differences, with personality traits and AI attitudes as covariates. Mediation analyses employed PROCESS Model 4 with bootstrap procedures for bias-corrected confidence intervals (Hayes, 2017).

**4.2.5.1.1 Manipulation checks.** Participants received two binary comprehension questions assessing accurate processing of framing instructions. Question 1 assessed prototypicality manipulation comprehension; Question 2 assessed emotional involvement manipulation comprehension. Both required correct responses for inclusion.

#### 4.2.5.2 Ethics

The Kansai University Graduate School of Psychology Ethics Review Committee approved all Study 2 procedures in August 2024 (#409). Anonymity was ensured through randomly generated identifiers with encrypted data transmission and storage. The research plan and all question items were included in the ethical consideration form.

## 4.3 Results

### 4.3.1 Replication and context-dependent effects of classification

Descriptive statistics are shown in Table 4. The two-factor structure replicated: exploratory factor analysis (Table 5) supported a two-factor solution (eigenvalues 14.89, 3.23; variance 42.3%), with Factor 1 capturing institutional disputes (patent infringement $\lambda = 0.973$, contractual violations $\lambda = 0.867$) and Factor 2 capturing interpersonal disputes (child custody $\lambda = 0.947$, child abuse $\lambda = 0.754$). Factor intercorrelation ($r = 0.791$) closely matched Study 1 ($r = 0.805$), confirming structural stability. Composite scores demonstrated systematic differentiation: institutional disputes (M = 2.67, SD = 0.89) versus interpersonal disputes (M = 3.12, SD = 0.84), $d = 0.52$, $F(1, 596) = 12.69$, $p < 0.001$, $\eta^2 = 0.021$.

Both factor means exceeded the midpoint (2.5), confirming domain-sensitive differentiation rather than categorical AI rejection. These findings demonstrate a stable psychological structure organizing preferences across independent samples.

### 4.3.2 Personality assessment

The TIPI-J (Oshio et al., 2012) demonstrated psychometric properties typical of ultra-brief personality measures. Exploratory factor analysis using five fixed factors revealed personality dimensions were defined through designated item pairs, although neuroticism and openness displayed moderate cross-loadings (Table 6). Confirmatory factor analysis yielded mixed fit indices (CFI = 0.789, TLI = 0.620, RMSEA = 0.186), reflecting psychometric limitations of ultra-brief scales.

Internal consistency ranged from adequate to modest ($\alpha = 0.41$ to 0.63; $\omega = 0.44$ to 0.72), aligning with two-item scale norms. Despite limitations, TIPI-J demonstrated meaningful relationships with criterion variables, providing a personality measurement suitable for experimental contexts with survey length constraints.

#### 4.3.2.1 Assessment of attitudes toward AI

The AI Expectations and Risk Perceptions Scales (Ota, 2023) demonstrated excellent psychometric properties ($\alpha = 0.912$ and 0.896, respectively) with clear unidimensional structures. AI Expectations averaged 29.18 (SD = 5.93), Risk Perceptions averaged 29.81 (SD = 5.48). Paired-samples $t$-test revealed marginally higher risk scores, $t(595) = -1.949$, $p = 0.052$, $d = 0.11$, indicating cautious optimism toward AI.

The moderate negative correlation ($r = -0.34$, $p < 0.001$) confirmed they represent related but distinct dimensions rather than opposite ends of a continuum.

### 4.3.3 Effects of experimental manipulation

Univariate ANOVA revealed a significant three-way interaction among emotional involvement, gender, and prototypicality, $F(1, 587) = 4.071$, $p = 0.044$, $\eta^2 = 0.007$ (Figure 2), providing evidence for context-dependent cognitive classification processes.

Under prototypical conditions, emotional involvement × gender interaction was significant, $F(1, 587) = 15.265$, $p < 0.001$, $\eta^2 = 0.025$. Men displayed higher AI acceptance under emotional framing (M = 110.24) than neutral framing (M = 118.15), $t(158) = 2.18$, $p = 0.031$, $d = 0.31$. Women demonstrated stronger human judgment preference under emotional framing (M = 129.34) than neutral framing (M = 111.09), $t(227) = -5.89$, $p < 0.001$, $d = 0.74$.

Under non-prototypical conditions, no significant interaction emerged, $F(1, 587) = 0.46$, $p = 0.499$, $\eta^2 = 0.003$, implying uniform perception that rare disputes require human judgment regardless of emotional context.

The complex interaction supports context-dependent classification. Women's strong human judgment preference under emotional framing ($d = 0.74$) indicates activation of classification schemes emphasizing empathy. Men's increased AI acceptance may reflect schemes prioritizing objective evaluation.

### 4.3.4 Moderating effects of individual differences and predictors of AI attitudes

Mixed-design ANCOVA confirmed the significant main effect of dispute type, $F(1, 596) = 12.69$, $p < 0.001$, $\eta^2 = 0.021$. Gender emerged as significant, $F(1, 586) = 17.25$, $p < 0.001$, $\eta^2 = 0.029$, with women consistently exhibiting stronger preference for human judgment (M = 123.67) than men (M = 115.23), $d = 0.35$.

The interaction between age and dispute type was significant, $F(1, 596) = 7.29$, $p = 0.007$, $\eta^2 = 0.012$, revealing that older participants demonstrated larger differences between dispute types. The interaction between extraversion and dispute type was also significant, $F(1, 596) = 9.48$, $p = 0.002$, $\eta^2 = 0.016$.

### 4.3.5 Effects of AI attitude

AI expectations emerged as the strongest predictor of acceptance, $F(1, 586) = 197.43$, $p < 0.001$, $\eta^2 = 0.252$, significantly exceeding the predictive power of personality traits and demographics. AI risk perceptions functioned as a significant negative predictor, $F(1, 586) = 25.25$, $p < 0.001$, $\eta^2 = 0.041$.

For AI expectations, agreeableness emerged as the strongest predictor, $F(1, 588) = 20.94$, $p < 0.001$, $\eta^2 = 0.034$. A significant gender × agreeableness interaction emerged, $F(1, 586) = 7.83$, $p = 0.005$, $\eta^2 = 0.013$.

### 4.3.6 Sensitivity analyses: validation of exclusion criteria

To evaluate whether stringent exclusion criteria influenced experimental effects, we conducted comparative analyses across three quality-control scenarios: (a) no exclusion ($N = 1{,}818$); (b) exclusion for failed attention checks only ($N = 1{,}075$); and (c) exclusion for both attention and manipulation-comprehension check failures ($N = 596$). The fully screened sample produced a statistically significant three-way interaction, $F(1, 1{,}067) = 6.276$, $p = 0.012$, partial $\eta^2 = 0.006$. The attention-check-only sample failed to yield significant three-way interaction, $F(1, 1{,}067) = 1.000$, $p > 0.05$, demonstrating that low-engagement respondents contributed error variance that obscured causal effects.

## 4.4 Discussion

### 4.4.1 Replication and contextual modulation, and robustness of dimensional structure

Study 2 successfully replicated the two-dimensional structure identified in Study 1, demonstrating stability across independent samples and measurement formats. The dimensional structure emerged consistently despite Study 2's inclusion of a neutral midpoint (five-point scale versus Study 1's four-point forced-choice format), suggesting that the interpersonal-institutional distinction reflects a robust psychological organization rather than measurement artifact. Factor intercorrelations remained highly consistent ($r = 0.791$ in Study 2 versus $r = 0.805$ in Study 1), and composite score differentiation ($d = 0.52$) confirmed systematic preference variation across dispute types.

Experimental manipulations produced theoretically meaningful effects, with contextual features systematically modulating acceptability judgments. The significant three-way interaction among emotional involvement, gender, and prototypicality ($\eta^2 = 0.007$) demonstrates that contextual effects operate conditionally rather than uniformly. Under prototypical conditions, emotional framing produced strong gender-differentiated effects: women showed substantially increased preference for human judgment ($d = 0.74$), while men showed modest increased AI acceptance ($d = 0.31$). Under non-prototypical conditions, both genders uniformly preferred human judgment regardless of emotional framing, suggesting that

TABLE 4 Descriptive statistics for each question item.

| Item | Mean | SD |
|---|---|---|
| Q01 Valuables were stolen from my neighbor | 2.520 | 1.026 |
| Q02 Motorists ran a red light and crashed into each other | 2.430 | 1.049 |
| Q03 One got into a fight and hurt someone else | 2.780 | 1.055 |
| Q04 One bought a product on the internet but did not receive it | 2.310 | 0.984 |
| Q05 One was slandered on the internet | 2.520 | 1.093 |
| Q06 A married couple had trouble and they filed for divorce | 3.230 | 1.048 |
| Q07 A public official received a bribe in connection with their duties | 2.420 | 1.064 |
| Q08 One bought goods from a seller in person, but the quality of the goods was lower than described at the time of purchase | 2.700 | 1.026 |
| Q09 One trespassed on someone else's land without permission | 2.520 | 1.036 |
| Q11 One drank alcohol and caused a commotion in the park | 2.540 | 1.006 |
| Q12 Someone was violent toward their child | 2.980 | 1.155 |
| Q13 One interfered with the progress of a trial in court | 2.550 | 1.039 |
| Q14 A group of drug traffickers was caught | 2.300 | 1.055 |
| Q15 One gave false testimony at a trial | 2.580 | 1.093 |
| Q16 Prostitution was conducted in an illegal brothel | 2.530 | 1.094 |
| Q17 A suspect was arrested for murdering another person | 2.810 | 1.138 |
| Q18 A person profited from copying another person's copyrighted work without permission | 2.290 | 1.008 |
| Q19 The advertisement of a product is different from the actual product | 2.420 | 0.985 |
| Q20 One got followed or harassed by a stalker | 2.800 | 1.153 |
| Q21 Someone forced somebody to repay a debt through illegal means | 2.570 | 1.086 |
| Q22 A worker's health was damaged due to overwork | 2.800 | 1.116 |
| Q23 One obstructed a police officer in the performance of their duties | 2.610 | 1.014 |
| Q24 One struck a person in a fight | 2.870 | 1.049 |
| Q25 One helped a person who was asked to commit suicide | 3.260 | 1.020 |
| Q26 One improperly used another company's patent or trademark | 2.330 | 1.025 |
| Q27 One drove a car after consuming alcohol | 2.270 | 1.066 |
| Q28 An underage person consumed alcohol | 2.470 | 1.013 |
| Q29 A company failed to comply with environmental laws and polluted the air or water | 2.410 | 1.067 |
| Q30 A company leaked the personal information of a customer | 2.360 | 1.042 |
| Q31 A company discriminated against women in terms of employment and wages | 2.770 | 1.111 |
| Q32 A municipality made a decision that violated an individual's human rights | 2.680 | 1.077 |
| Q33 A newly built house was defective and I sought repairs or compensation | 2.450 | 1.046 |
| Q35 My supervisor or co-worker made sexually explicit comments or actions | 2.930 | 1.108 |
| Q36 A building collapsed and people were injured or killed | 2.700 | 1.113 |
| Q37 An employer fired a person for illegal reasons | 2.710 | 1.053 |
| Q38 One was involved in an online scam | 2.320 | 1.022 |
| Q39 A violent incident occurred in a public place and there were victims | 2.660 | 1.073 |
| Q40 One failed to fulfill its contractual obligations | 2.430 | 0.991 |
| Q41 A restaurant was caught violating food safety laws | 2.390 | 1.014 |
| Q42 A fraudster provided false information to obtain a loan | 2.410 | 1.050 |
| Q43 A dispute over custody of children arose between a divorcing couple | 3.240 | 1.073 |
| Q44 A company was alleged to have been involved with antisocial forces | 2.520 | 1.047 |
| Q45 A robbery was committed with a weapon | 2.440 | 1.099 |
| Q46 The accused sexually assaulted a victim | 2.920 | 1.172 |

*(Continued)*

TABLE 4 (Continued)

| Item | Mean | SD |
|---|---|---|
| Q47 Someone trafficked in large quantities of narcotics | 2.230 | 1.068 |
| Q48 A group committed a major fraud | 2.390 | 1.106 |

The table reports the mean values and standard deviations (SD) for each of the 48 question items assessed on a 5-point scale ($N = 596$). The items represent a broad range of interpersonal, social, and organizational transgressions, such as theft, violence, fraud, discrimination, and corporate misconduct. These descriptive statistics provide a baseline understanding of participants' evaluations of each scenario. $N = 596$ 5-point scales.

perceived novelty or complexity may override other classification-relevant features.

These patterns confirm that acceptability judgments respond to contextual manipulation in theoretically predicted ways. However, Study 2 manipulated features theorized to influence how disputes are mentally represented without directly measuring classification processes. Observed effects are consistent with classification-based interpretation but do not definitively establish classification as the mediating mechanism. We return to this interpretive issue in the General Discussion, where we consider alternative psychological processes and propose methodological approaches for stronger causal inference.

Several findings prove robust across both studies: (1) acceptability judgments vary systematically with dispute content in ways that align with theoretical distinctions from moral psychology and construal level theory; (2) contextual features systematically modulate judgments in predictable directions; (3) individual differences—particularly AI-specific attitudes—exert substantial direct effects and moderate contextual influences; (4) domain-specific attitudes constitute the strongest proximal predictor of acceptance, substantially exceeding the predictive power of personality traits or demographics.

### 4.4.2 Limitations and boundary conditions

The 67.2% exclusion rate raises questions about generalizability. The retained sample represents individuals capable of processing experimental instructions. Whether findings generalize to less deliberative citizens remains uncertain.

The vignette-based methodology, while facilitating experimental control, may not capture the richness of authentic legal disputes. The extent to which observed patterns would replicate with actual case materials remains an empirical question.

Cultural specificity of findings within the Japanese legal context requires caution. Gender effects may reflect culturally specific socialization patterns. Cross-cultural replication is necessary.

To assess potential selection effects from exclusion procedures, we compared demographic characteristics between retained ($N = 596$) and excluded ($N = 1,222$) participants. Retained participants were older on average ($M_{age} = 55.45$) than excluded participants ($M_{age} = 51.57$). Gender composition differed significantly, $\chi^2(1) = 16.80$, $p < 0.001$, Cramér's $V = 0.096$, with women overrepresented among retained (58.6%) versus excluded (48.2%) participants. Occupational distributions also differed significantly, $\chi^2(11) = 36.48$, $p < 0.001$, Cramér's $V = 0.142$, indicating modest employment composition shifts. Household income distributions showed a smaller but significant difference, $\chi^2(14) = 23.72$, $p = 0.050$, Cramér's $V = 0.114$. These results suggest the analytical sample modestly overrepresents older, female, and certain occupational categories. Although digital literacy and AI familiarity were not directly measured, these demographic differences indicate potential selection bias, warranting cautious interpretation of generalizability.

### 4.4.3 Implications and contribution

Study 2 advances understanding by demonstrating that acceptability judgments are systematically influenced by contextual features of disputes beyond stable individual differences. The replication of dimensional structure across independent samples provides convergent support for the stability of observed patterns. The experimental evidence for contextual modulation—particularly the interaction between prototypicality and emotional involvement—indicates that acceptance is not merely a function of fixed attitudes but reflects context-dependent evaluation processes.

Several implications emerge for theory and practice. Theoretically, the findings suggest that technology acceptance models may benefit from incorporating contextual features that vary across decision situations. The person-by-situation interactions observed in Study 2 highlight the limitations of models emphasizing either individual differences or situational features in isolation. Integrative frameworks acknowledging multiple pathways—direct effects of attitudes, moderation of contextual features by dispositions, and reciprocal influences between situational construal and evaluative judgments—may be necessary to fully capture acceptance dynamics.

Practically, the findings provide guidance for phased implementation strategies. High-consensus domains (institutional disputes, prototypical cases) may constitute optimal starting points for AI deployment, where public acceptance is relatively higher and less variable. For interpersonal disputes and atypical cases, maintaining robust human oversight and transparently communicating system limitations may be essential for legitimacy. The prominence of AI-specific attitudes suggests that targeted educational initiatives addressing capabilities and limitations may prove more effective than generic awareness campaigns.

The gender-differentiated processing patterns observed in Study 2 suggest potential value in audience segmentation for communication strategies. However, we caution against overgeneralization—gender effects may reflect cultural factors specific to Japan and should not be assumed to generalize without empirical replication across diverse populations.

Most fundamentally, Study 2 demonstrates that the content of legal disputes—not merely individual attitudes or technology capabilities—systematically shapes public acceptance. This situational variability has important implications for implementation planning, suggesting that deployment strategies should be tailored to dispute characteristics rather than applied uniformly across judicial domains.

## 5 General discussion

The present research investigated whether psychological features of legal dispute content—beyond individual differences—shape public acceptance of AI adjudication. Across two studies with Japanese

TABLE 5 Factor analysis for each question item in Study 2.

| Item | Factor1 | Factor2 |
|---|---|---|
| Q26 One improperly used another company's patent or trademark | 0.973 | |
| Q47 Someone trafficked in large quantities of narcotics | 0.933 | |
| Q18 A person profited from copying another person's copyrighted work without permission | 0.923 | |
| Q38 One was involved in an online scam | 0.912 | |
| Q42 A fraudster provided false information to obtain a loan | 0.866 | |
| Q14 A group of drug traffickers was caught | 0.866 | |
| Q30 A company leaked the personal information of a customer | 0.844 | |
| Q48 A group committed a major fraud | 0.831 | |
| Q27 One drove a car after consuming alcohol | 0.826 | |
| Q45 A robbery was committed with a weapon | 0.819 | |
| Q41 A restaurant was caught violating food safety laws | 0.810 | |
| Q07 A public official received a bribe in connection with their duties | 0.759 | |
| Q04 One bought a product on the internet but did not receive it | 0.750 | |
| Q19 The advertisement of a product is different from the actual product | 0.739 | |
| Q40 One failed to fulfill its contractual obligations | 0.734 | |
| Q29 A company failed to comply with environmental laws and polluted the air or water | 0.721 | |
| Q33 A newly built house was defective and I sought repairs or compensation | 0.700 | |
| Q02 Motorists ran a red light and crashed into each other | 0.686 | |
| Q16 Prostitution was conducted in an illegal brothel | 0.682 | |
| Q01 Valuables were stolen from my neighbor | 0.661 | |
| Q09 One trespassed on someone else's land without permission | 0.652 | |
| Q28 An underage person consumed alcohol | 0.624 | |
| Q44 A company was alleged to have been involved with antisocial forces | 0.600 | |
| Q21 Someone forced somebody to repay a debt through illegal means | 0.595 | |
| Q13 One interfered with the progress of a trial in court | 0.572 | |
| Q15 One gave false testimony at a trial | 0.545 | |
| Q23 One obstructed a police officer in the performance of their duties | 0.526 | |
| Q11 One drank alcohol and caused a commotion in the park | 0.519 | |
| Q32 A municipality made a decision that violated an individual's human rights | 0.463 | |
| Q36 A building collapsed and people were injured or killed | 0.458 | |
| Q08 One bought goods from a seller in person, but the quality of the goods was lower than described at the time of purchase | 0.427 | |
| Q05 One was slandered on the internet | 0.425 | |
| Q43 A dispute over custody of children arose between a divorcing couple | | 0.947 |
| Q25 One helped a person who was asked to commit suicide | | 0.880 |
| Q06 A married couple had trouble and they filed for divorce | | 0.879 |
| Q12 Someone was violent toward their child | | 0.754 |
| Q35 My supervisor or co-worker made sexually explicit comments or actions | | 0.738 |
| Q46 The accused sexually assaulted a victim | | 0.681 |
| Q20 One got followed or harassed by a stalker | | 0.644 |
| Q03 One got into a fight and hurt someone else | | 0.579 |
| Q22 A worker's health was damaged due to overwork | | 0.567 |
| Q31 A company discriminated against women in terms of employment and wages | | 0.567 |
| Q24 One struck a person in a fight | | 0.560 |
| Q17 A suspect was arrested for murdering another person | | 0.498 |

*(Continued)*

TABLE 5 (Continued)

| Item | Factor1 | Factor2 |
|---|---|---|
| Q37 An employer fired a person for illegal reasons | | 0.483 |
| Q39 A violent incident occurred in a public place and there were victims | | 0.461 |

The table displays factor loadings from an exploratory factor analysis conducted on 48 question items in Study 2 ($N$ = 596). Two factors were extracted. Factor 1 captured items primarily related to corporate and societal misconduct (e.g., fraud, data leakage, intellectual property violations, environmental non-compliance), whereas Factor 2 represented items concerning interpersonal and violent behaviors (e.g., assault, child abuse, homicide, sexual misconduct). Only loadings greater than 0.40 are reported.

TABLE 6 Pattern matrix of TIPI-J.

| | | Factor | | | | |
|---|---|---|---|---|---|---|
| Big-Five trait | Item | 1 | 2 | 3 | 4 | 5 |
| Neuroticism | Calm, emotionally stable (R) | −0.782 | | | | |
| Neuroticism | Anxious, easily upset | −0.763 | | 0.329 | | |
| Agreeableness | Critical, quarrelsome (R) | 0.306 | | | | |
| Extraversion | Extraverted, enthusiastic | | 0.855 | | | |
| Extraversion | Reserved, quiet (R) | | 0.673 | | | |
| Agreeableness | Sympathetic, warm | | | 0.721 | | |
| Conscientiousness | Dependable, self-disciplined | | | 0.436 | | |
| Openness to experience | Conventional, uncreative (R) | | | | 0.712 | 0.342 |
| Conscientiousness | Disorganized, careless (R) | | | | 0.481 | |
| Openness to experience | Open to new experiences, complex | | | | | 0.809 |

The table presents the factor loadings for the 10 items of the Japanese version of the Ten Item Personality Inventory (TIPI-J) across five designated factors corresponding to the Big-Five personality traits ($N$ = 596). A promax rotation was applied. Items marked with "(R)" represent reverse-coded items. Factor loadings greater than |0.30| are displayed.

participants, we observed robust evidence that acceptability judgments vary predictably with theoretically meaningful dispute characteristics. Study 1 revealed a two-dimensional structure distinguishing interpersonal-relational disputes (where human adjudicators were strongly preferred) from institutional-procedural disputes (where AI acceptance was comparatively higher). Study 2 replicated this structure and demonstrated that experimentally manipulated contextual features—emotional involvement and prototypicality—modulated acceptability judgments, with effects moderated by AI-specific attitudes and gender.

These findings suggest dispute content constitutes an overlooked dimension in AI acceptance research. While prior work has documented individual differences, our results indicate situational features—psychological characteristics of decision contexts—contribute substantially to preferences for AI versus human adjudication.

Table 7 summarizes core findings across both studies, including the dimensional structure identified in Study 1, its replication and experimental extensions in Study 2, and the roles of contextual features and individual differences in shaping acceptability judgments. This compact reference highlights convergent patterns, boundary conditions, and theoretical relevance, facilitating integration with the detailed implications discussed below.

## 5.1 Theoretical implications

### 5.1.1 Situating dispute content in AI acceptance models

Current models of AI acceptance—including the Technology Acceptance Model (Davis, 1989) and Computers Are Social Actors framework (Nass and Moon, 2000)—predominantly emphasize stable individual differences. Our findings suggest these models may benefit from incorporating contextual features varying across decision situations. The patterns we observed align with theoretical distinctions in moral psychology and construal level theory.

The interpersonal-relational versus institutional-procedural distinction mirrors established frameworks. Social domain theory (Turiel, 1983) distinguishes moral transgressions from conventional violations, positing these domains are cognitively represented differently. Construal level theory (Trope and Liberman, 2010) proposes psychologically proximal situations evoke concrete, contextualized processing, whereas distal situations promote abstract, rule-based processing. Our findings are consistent with disputes involving interpersonal harm being construed as requiring empathetic human judgment, whereas rule-based institutional disputes may be viewed as amenable to algorithmic processing.

The framework extends technology acceptance models by incorporating psychological features of decision contexts as sources of variance, aligning with person-situation interactions in social judgment (Mischel and Shoda, 1995) and domain-specificity in moral cognition (Turiel, 1983).

### 5.1.2 Cognitive mechanisms: interpretation and alternatives

We interpreted the observed dimensional structure through cognitive categorization—that individuals classify legal disputes into psychologically meaningful types, differentially affording AI versus human adjudication. This interpretation is theoretically motivated and consistent with category-based reasoning in legal contexts (Smith et

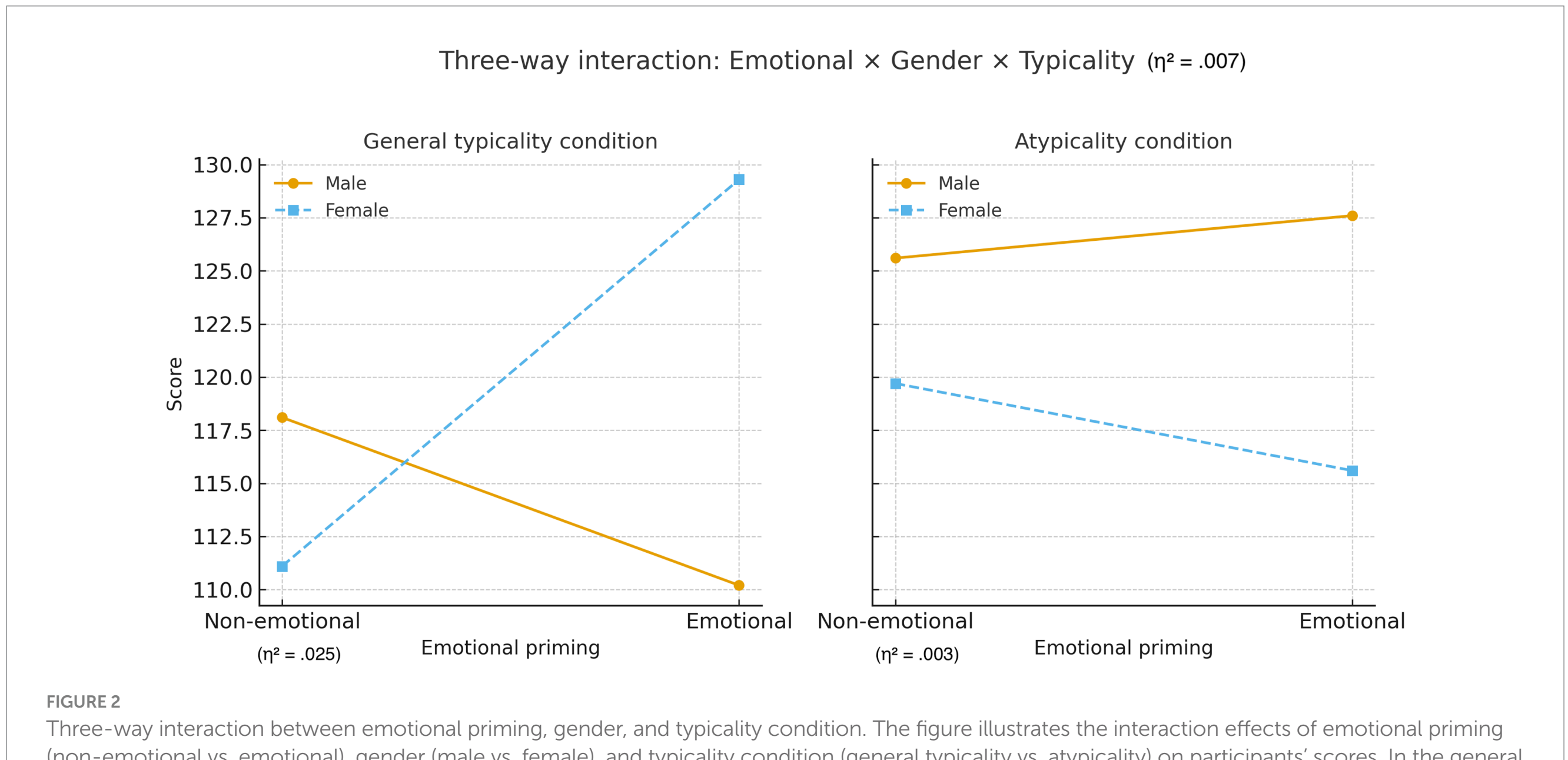

FIGURE 2
Three-way interaction between emotional priming, gender, and typicality condition. The figure illustrates the interaction effects of emotional priming (non-emotional vs. emotional), gender (male vs. female), and typicality condition (general typicality vs. atypicality) on participants' scores. In the general typicality condition (left panel), males scored higher in the non-emotional priming condition, whereas females showed markedly higher scores in the emotional priming condition, indicating a crossover interaction ($\eta^2$ = 0.025). In the atypicality condition (right panel), scores were relatively stable, with males showing a slight increase and females a slight decrease from non-emotional to emotional priming ($\eta^2$ = 0.003). The overall three-way interaction: $F(1, 587) = 4.071$, $p = 0.044$, $\eta^2 = 0.007$.

TABLE 7 Summary of main findings across studies.

| Aspect | Study 1 | Study 2 |
|---|---|---|
| Dimensional structure | 2 factors (institutional/ interpersonal) | Replicated |
| Scale format | 4-point forced | 5-point Likert |
| Manipulation effects | — | Significant |
| Exclusion rate | 31% | 67% |
| Robustness checks | EFA | EFA + sensitivity |

This table summarizes the main empirical results, including the dimensional structure of dispute classification identified in Study 1, its replication and experimental extensions in Study 2, and effects of contextual features and individual differences on acceptability judgments. It highlights convergent patterns, boundary conditions, and theoretically relevant variables discussed in the General Discussion.

al., 1992) and dual-process models of moral judgment (Greene et al., 2001).

However, three caveats apply. First, our studies assessed acceptability judgments but did not directly measure cognitive classification processes. The dimensional structure represents a descriptive summary of response patterns, not direct evidence for classification as a psychological mechanism.

Second, alternative processes may generate similar patterns. Emotional responses to dispute content may drive preferences independently of explicit categorization. Moral intuitions or perceived fairness considerations may similarly shape judgments. Our studies cannot adjudicate between these mechanisms.

Third, the causal direction between dispute representation and acceptability remains uncertain. While we framed psychological features as influencing preferences, pre-existing preferences may shape how disputes are construed. Disentangling these pathways would require experimental manipulation of classification itself.

Despite limitations, convergence of findings across exploratory and experimental approaches, replication across independent samples, and systematic modulation by contextual manipulations suggest psychological features of dispute content play a meaningful role in shaping AI acceptance.

### 5.1.3 Individual differences and mediation pathways

AI-specific attitudes emerged as the strongest predictor of acceptance ($\eta^2$ = 0.252), substantially exceeding demographic factors and personality traits. This confirms domain-specific evaluative orientations exert greater influence than stable individual characteristics, aligning with technology acceptance research (Venkatesh and Davis, 2000; Lee, 2018).

Gender moderating effects revealed important boundary conditions. For men, agreeableness strongly predicted AI expectations ($b = 1.40$, $p < 0.001$), whereas this relationship was weaker for women ($b = 0.41$, $p = 0.062$). Under prototypical conditions, emotional involvement exhibited divergent gender effects: women displayed increased human judgment preference under emotional framing ($d = 0.74$), while men demonstrated increased AI acceptance.

These patterns may reflect culturally specific mechanisms within the Japanese context. Cross-cultural replication is essential. The prominence of AI-specific attitudes over personality traits has practical implications: educational initiatives targeting technological capabilities may prove more effective than broad-based approaches.

### 5.1.4 Boundary conditions and context dependency

Study 2 revealed that contextual effects were moderated by prototypicality. Under prototypical conditions, emotional involvement produced gender-differentiated effects. Under non-prototypical conditions, both genders uniformly preferred human judgment regardless of emotional framing.

This hierarchical organization suggests that individuals employ structured evaluation strategies where certain features (prototypicality) determine whether other features (emotional involvement) influence judgments. The absence of interaction effects under non-prototypical conditions indicates perceived novelty or complexity may constitute a threshold condition overriding other considerations.

### 5.1.5 Alternative psychological explanations

The institutional–interpersonal structure may overlap with established constructs, particularly moral intuitions and fairness perceptions. Interpersonal disputes may activate care-based moral foundations, increasing demands for human empathy, whereas institutional disputes may invoke authority norms, rendering algorithmic consistency acceptable. Procedural justice theory similarly emphasizes voice, respect, and neutrality as legitimacy determinants; participants may reject AI in emotionally laden disputes due to heightened expectations of interpersonal recognition rather than categorical classification per se. However, these accounts do not fully explain the consistent two-dimensional structure across formats and samples. Moral intuitions and fairness concerns likely contribute to, rather than replace, the classification process. Future research integrating measures of moral foundations, fairness sensitivity, and classification tendencies would clarify their relative explanatory roles and causal ordering.

## 5.2 Methodological considerations and limitations

### 5.2.1 Causal inference and psychological mechanisms

The fundamental limitation concerns causal inference regarding the relationship between cognitive classification and AI acceptance. Although we interpret findings through classification theory, the research does not definitively establish that classification processes causally produce acceptance judgments. Multiple interpretive frameworks remain plausible, and evidence supporting classification-based interpretation, while consistent, remains indirect.

Three causal models warrant consideration. First, the classification-causes-acceptance model proposes individuals first cognitively organize disputes into meaningful categories (interpersonal versus institutional), which subsequently shape judgments about appropriate adjudicative processes. This represents our preferred interpretation, aligned with categorization theory (Rosch, 1975; Murphy, 2004) and construal level theory (Trope and Liberman, 2010).

Second, the common-cause model proposes both classification and acceptance reflect shared underlying processes—particularly emotional responses or moral intuitions—without classification causally producing acceptance. Emotional reactions to dispute content might simultaneously influence both how disputes are categorized and how adjudicative technologies are evaluated, producing observed correlations without causal precedence (Chaiken and Trope, 1999).

Third, the reverse-causality model proposes pre-existing AI attitudes influence how disputes are classified, rather than classification influencing acceptance. Citizens with strong pro-AI attitudes might mentally represent diverse disputes as rule-based institutional problems amenable to algorithmic processing, whereas those with pro-human attitudes might construe the same disputes as interpersonally complex situations requiring human judgment.

The present studies cannot definitively adjudicate among these models. Study 1's exploratory factor analysis revealed dimensional structure but assessed classification and acceptance simultaneously, precluding temporal ordering. Study 2's experimental manipulations demonstrated contextual features modulate acceptance judgments, but manipulated theorized antecedents of classification (emotional involvement, prototypicality) rather than directly manipulating classification itself or measuring classification processes independently.

Several patterns provide circumstantial support for classification-based interpretation. The dimensional structure proved robust across independent samples, measurement formats, and analytical approaches, suggesting stable psychological organization. Experimental effects aligned with theoretical predictions: emotional involvement increased human preference particularly for women, prototypicality moderated contextual effects, and effects operated conditionally based on individual characteristics. These patterns are consistent with classification mediating between contextual features and acceptance.

However, alternative mechanisms could generate similar patterns. Emotional framing may activate empathetic responses through affective pathways independent of explicit categorization. Prototypicality framing may influence perceived decision complexity, risk, or stakes through mechanisms unrelated to domain classification. Gender differences may reflect socialized differences in empathy, risk perception, or trust rather than differences in classification processes.

Establishing genuine causal precedence would require different research designs. Multi-phase longitudinal studies where participants first classify disputes then evaluate AI suitability following temporal delay would enable examination of whether classification predicts later acceptance changes. Experimental designs manipulating classification cues independently—through priming emphasizing institutional versus relational features—before presenting acceptance judgments would clarify whether induced classification shifts produce corresponding evaluation changes. Process-tracing methodologies including think-aloud protocols, reaction time paradigms, and neuroimaging approaches would provide more direct evidence regarding temporal sequencing and mediating mechanisms.

Despite limitations, the research makes important theoretical contributions. We demonstrate acceptability judgments are not merely functions of individual differences or general technology attitudes but vary systematically with psychological characteristics of decision contexts. This situational specificity extends technology acceptance models in theoretically meaningful directions. Whether classification constitutes the specific mechanism producing context-dependent acceptance, or whether classification and acceptance co-vary through shared emotional and moral processes, the practical implication remains: AI acceptance in judicial contexts depends fundamentally on what type of decision is being delegated, not merely on who evaluates the technology or what capabilities it possesses.

### 5.2.2 Measurement of cognitive processes and construct validity

The primary methodological limitation concerns construct validity and the inferential gap between observed response patterns and underlying psychological processes. Across both studies, we inferred dimensional structure from acceptability ratings but did not independently assess cognitive classification processes. This creates potential circularity: we interpret dimensional structure as evidence for classification, but dimensions derive from the same acceptability judgments they purport to explain.

This limitation manifests in multiple ways. First, dimensional structure emerged from exploratory factor analysis in Study 1, meaning that the resulting dimensions represent descriptive summaries of response patterns rather than direct evidence for underlying mechanisms. Alternative mechanisms—including emotional responses independent of explicit categorization, moral intuitions activating domain-specific processing, or fairness considerations influencing procedural preferences—could generate similar dimensional patterns without requiring classification-based mediation.

Second, our stimulus materials consisted of brief, researcher-constructed vignettes designed to sample diverse legal domains while maintaining experimental control. This facilitated broad coverage and enabled identification of systematic preference patterns, but may limit ecological validity compared to authentic legal materials. Actual legal disputes involve complex factual patterns, procedural histories, and contextual nuances not fully captured in single-sentence descriptions. Whether observed dimensional structure would replicate with actual case materials from court databases or judicial opinions remains an empirical question.

Third, Study 1 employed a four-point forced-choice response format enhancing discrimination but precluding expressing equivalence between AI and human adjudicators. While theoretically motivated—to counter documented tendencies toward middle-category selection among Japanese respondents (Chen et al., 1995; Harzing, 2006)—the forced-choice format may have artificially inflated dimensional differentiation. Study 2 addressed this by including a neutral midpoint, and replication of dimensional structure provides reassurance. However, systematic comparison across diverse response formats would strengthen confidence in the psychological reality of observed structure.

While exploratory factor analysis is methodologically appropriate for uncovering latent structure, and confirmatory analysis in Study 2 provided replication, resulting dimensions remain fundamentally descriptive rather than explanatory. We interpret findings through categorization and construal level theories, but this represents theoretical inference from observed patterns rather than direct empirical demonstration.

Promising methodological directions include: (1) direct classification tasks requiring participants to explicitly sort disputes before evaluating adjudicator appropriateness, (2) reaction time paradigms assessing whether classification-consistent judgments show faster latencies, (3) think-aloud protocols capturing real-time verbalization, (4) eye-tracking studies examining attention allocation, and (5) neuroimaging approaches identifying neural signatures of categorization versus evaluation. Such methods would provide stronger causal inference regarding classification's role in shaping AI acceptance.

#### 5.2.3 Stimulus materials and ecological validity

Our vignettes consisted of brief, researcher-constructed scenarios designed to sample diverse legal domains while controlling extraneous variables. This facilitated experimental control but may limit generalizability to real-world legal contexts. Actual disputes involve complex factual patterns and procedural histories not fully captured in one-sentence descriptions.

Enhancing ecological validity should be a priority. Potential approaches include using redacted case summaries from court databases, presenting excerpts from judicial opinions, employing video-recorded presentations, or conducting field studies with actual litigants or legal professionals.

#### 5.2.4 Sample characteristics and cross-cultural generalizability

Both studies employed Japanese online panel samples. High exclusion rates (31% in Study 1, 67% in Study 2) warrant caution. Online panels may not represent the broader population, particularly regarding digital literacy and AI familiarity.

Cultural factors specific to Japan—including collectivism, hierarchical social structures, and distinct legal culture (Kahan, 2015)—may influence dimensional structure. The interpersonal-institutional distinction may reflect culturally specific construals not generalizing to Western adversarial systems or other legal traditions. Cross-cultural replication is necessary to assess generalizability.

#### 5.2.5 Response scale decisions

Study 1 employed a 4-point forced-choice scale without a midpoint, enhancing discrimination but eliminating the ability to express equivalence between AI and human adjudicators. While Study 2 addressed this by including a midpoint, and key findings replicated, the forced-choice format may have artificially inflated dimensional differentiation in Study 1.

The theoretical concern is that removing the midpoint may have compelled participants to express preferences they would not naturally hold, creating a dimensional structure reflecting measurement artifact rather than genuine psychological organization. However, replication of dimensional structure in Study 2 (which included a midpoint) and conceptual alignment with established theoretical frameworks (social domain theory, construal level theory) provide reassurance the observed structure reflects more than measurement artifact.

Future research should systematically compare response formats—forced-choice versus midpoint-inclusive scales, different numbers of options, magnitude estimation—to assess robustness of dimensional structure across measurement approaches.

#### 5.2.6 Alternative explanations

Several alternative explanations warrant consideration. Observed dimensional structure might reflect familiarity rather than psychological categorization. Perceived stakes may vary systematically across dispute types. Trust in algorithmic systems may vary across domains independently of categorization processes. Disentangling these requires additional research designs.

### 5.3 Practical implications

#### 5.3.1 Phased implementation strategies

Our findings suggest AI deployment should begin with institutional-procedural disputes where acceptance is higher and variability lower. Traffic violations, regulatory compliance, and procedural standardization represent promising starting points. Interpersonal disputes involving family relationships, violent crimes, and child welfare should prioritize human judgment with AI serving supportive roles.

### 5.3.2 Communication strategies

The prominence of AI-specific attitudes as predictors suggests targeted educational initiatives may be more effective than generic campaigns. Communication efforts should address specific capabilities and limitations of AI systems. Transparency about algorithmic decision-making processes, error rates, and oversight mechanisms may help calibrate public expectations. Gender-differentiated effects suggest potential value in audience segmentation. However, we caution against overgeneralization—gender differences may reflect cultural factors specific to Japan.

### 5.3.3 Institutional design considerations

Our findings highlight the importance of hybrid systems combining algorithmic efficiency with human oversight. For institutional disputes, AI systems might provide preliminary assessments while preserving human authority for final decisions. For interpersonal disputes, AI might support information organization without replacing human judgment on substantive issues.

Procedural justice considerations remain paramount. Citizens' acceptance depends on perceptions of procedural fairness, voice, and respectful treatment (Tyler, 2006). Implementation frameworks should incorporate mechanisms for appeal, explanation, and human review.

## 5.4 Ecological validity and scenario abstraction

The use of single-sentence vignettes provides experimental control but abstracts from real-world legal complexity. Minimal scenarios may encourage prototype-based reasoning, potentially amplifying categorical distinctions that richer informational contexts would attenuate. Actual legal settings involve layered narratives with procedural history, evidentiary uncertainty, institutional context, and emotional nuance—factors that may interact with adjudicator preferences in ways brief descriptions cannot capture. Future studies should employ redacted case summaries, multi-paragraph narratives, or simulated legal documents to assess whether the institutional–interpersonal distinction persists under cognitively demanding, information-rich conditions or merely reflects abstraction-induced heuristics. Such designs would also examine how contextual overload, ambiguity, and conflicting cues moderate AI acceptance, thereby enhancing the external validity of classification-based models.

## 5.5 Future research directions

Although these studies reveal a robust two-dimensional structure in public acceptance judgments, the underlying cognitive processes remain inferred. Future research should shift from structural identification to process-level validation through three approaches. First, reaction time paradigms could test whether institutional and interpersonal disputes produce distinct latency patterns during adjudicator assignment, with faster responses to prototypical pairings indicating automatic classification. Second, mouse-tracking or eye-tracking could capture cognitive dynamics, revealing whether attention allocation varies by dispute type. Third, experimental manipulation of dispute features—emotional salience, precedent clarity, or relational proximity—could provide causal evidence that altering these dimensions shifts classification patterns. These methods would enable transition from dimensional description to mechanistic explanation in legal–technological contexts.

Several promising avenues emerge. First, direct manipulation of classification frameworks would allow stronger causal inference regarding cognitive mechanisms. Second, comparative research across legal cultures could assess whether the interpersonal-institutional distinction is culture-general. Third, longitudinal studies tracking acceptability as AI legal tools are deployed could reveal how experience shapes categorization and evaluation. Fourth, research should examine acceptability for hybrid human-AI systems. Fifth, research integrating measures of decision quality is essential. Finally, experimental research employing process-tracing methodologies could provide more direct evidence regarding cognitive mechanisms.

# 6 Conclusion

This research provides initial evidence that psychological features of legal dispute content—beyond individual differences—shape public acceptance of AI adjudication. The robust dimensional structure observed across two independent samples suggests individuals differentiate between dispute types in ways systematically influencing their procedural preferences.

The findings contribute to technology acceptance theory by demonstrating acceptance is not merely a function of who evaluates technology or what capabilities it possesses, but also what type of decision is being delegated. This situational specificity has important implications for implementation strategies.

As legal systems increasingly consider AI deployment, understanding how dispute characteristics shape public attitudes will be essential for designing technologies aligning with societal values. The interpersonal-institutional distinction provides one empirically grounded framework for anticipating resistance and identifying opportunities for acceptance-enhancing design.

Most fundamentally, our findings highlight public acceptance of AI in morally significant domains like law depends not only on demonstrating technical capability but also on aligning technological applications with citizens' intuitive understandings of what different problem types require. Technology acceptance in these contexts is less about overcoming resistance than achieving fit between technological affordances and psychological construals of decision tasks.

# Data availability statement

The raw data supporting the conclusions of this article will be made available by the authors, without undue reservation.

## Ethics statement

The studies involving humans were approved by the Ethics Committee of the Graduate School of Psychology, Kansai University. The studies were conducted in accordance with the local legislation and institutional requirements. The participants provided their written informed consent to participate in this study.

## Author contributions

MF: Conceptualization, Data curation, Formal analysis, Investigation, Methodology, Project administration, Visualization, Writing – original draft, Writing – review & editing. EW: Conceptualization, Funding acquisition, Investigation, Project administration, Writing – review & editing.

## Funding

The author(s) declared that financial support was received for this work and/or its publication. The authors received partial financial support for the research of this article from JSPS KAKENHI Grant Number JP22K03022.

## Acknowledgments

The authors thank the organizers and the participants of the Asian Law and Society Association conference in December 2023 for their insightful feedback on our research. The data analysis for Study 2, the drafting, and the revising of the manuscript were conducted during the first author's research-dedicated period at Kansai University, thus representing partial outcomes.

## Conflict of interest

The author(s) declared that this work was conducted in the absence of any commercial or financial relationships that could be construed as a potential conflict of interest.

## Generative AI statement

The author(s) declared that Generative AI was used in the creation of this manuscript. The author used AI while drafting the manuscript. According to the AI use policy of the journal, all prompts and outputs with the names of the AI models can be viewed through the following links: https://osf.io/kevqf/.

Any alternative text (alt text) provided alongside figures in this article has been generated by Frontiers with the support of artificial intelligence and reasonable efforts have been made to ensure accuracy, including review by the authors wherever possible. If you identify any issues, please contact us.

## Publisher's note

All claims expressed in this article are solely those of the authors and do not necessarily represent those of their affiliated organizations, or those of the publisher, the editors and the reviewers. Any product that may be evaluated in this article, or claim that may be made by its manufacturer, is not guaranteed or endorsed by the publisher.